\documentclass[11pt,fullpage]{article}
\usepackage{amsmath,graphicx,amsfonts,amssymb,sansseriftitles,natbib,bm}
\usepackage[title]{appendix}
\usepackage{algorithm,algorithmic}
\usepackage{caption}
\usepackage[section]{placeins}
\usepackage{lscape}
\usepackage{xcolor}
\usepackage[colorlinks=true,citecolor=blue,linkcolor=blue,urlcolor=blue]{hyperref}
\usepackage{subfigure}
\usepackage{setspace}
\usepackage{ragged2e}
\usepackage{multirow,booktabs}

\input epsf

\newcommand{\bi}{\begin{itemize}}
\newcommand{\ei}{\end{itemize}}
\newcommand{\be}{\begin{enumerate}}
\newcommand{\ee}{\end{enumerate}}

\newcommand {\bea}{\begin{eqnarray}}
\newcommand {\eea}{\end{eqnarray}}
\newcommand {\Pb}{\text{P}}
\newcommand {\Ex}{\text{E}}

\newcommand {\bs}{{\bf s}}

\newcommand {\ba}{{\bf a}}
\newcommand {\bI}{{\bf I}}

\newcommand {\bnu}{{\boldsymbol \nu}}

\newcommand {\bSigma}{{\bsym \Sigma}}
\newcommand {\bSigmaOld}{{\bsym \Sigma}_{\mbox{\tiny old}}}

\newcommand {\bSigmaNew}{{\bsym \Sigma}_{\mbox{\tiny new}}}

\newcommand {\pturn}{p^{\mbox{\tiny turn}}}

\newcommand {\sA}{s_{\mbox{\tiny A}}}
\newcommand {\sB}{s_{\mbox{\tiny B}}}

\newcommand {\snewB}{s_{\mbox{\tiny B}}^{\mbox{\tiny new}}}

\newcommand {\VNS}{V_{\mbox{\tiny NS}}}
\newcommand {\VA}{J^{\mbox{\tiny A-BR}}}

\newcommand {\piNS}{\pi^{\mbox{\tiny NS}}}
\newcommand {\piNSA}{\pi_{\mbox{\tiny A}}^{\mbox{\tiny NS}}}
\newcommand {\piNSB}{\pi_{\mbox{\tiny B}}^{\mbox{\tiny NS}}}
\newcommand {\piBRA}{\pi_{\mbox{\tiny A}}^{\mbox{\tiny BR}}}

\newcommand {\piA}{\pi_{\mbox{\tiny A}}}
\newcommand {\piB}{\pi_{\mbox{\tiny B}}}
\newcommand {\pA}{p_{\mbox{\tiny A}}}
\newcommand {\pB}{p_{\mbox{\tiny B}}}
\newcommand {\PA}{P_{\mbox{\tiny A}}}

\newcommand {\VVA}{V_{\mbox{\tiny A}}}
\newcommand {\VVB}{V_{\mbox{\tiny B}}}

\newcommand {\VNST}{V_{\mbox{\tiny NS}}}

\newcommand {\lcom}{l_{\mbox{\tiny com}}}
\newcommand {\JpiApib}{J^{{\tiny \pi_{\mbox A}, \pi_{\mbox B}}}}
\newcommand {\JpisApibs}{J^{{\tiny \pi_{\mbox A}^*, \pi_{\mbox B}^*}}}

\newcommand {\JpiApibs}{J^{{\tiny \pi_{\mbox A}, \pi_{\mbox B}^*}}}
\newcommand {\JpisApib}{J^{{\tiny \pi_{\mbox A}^*, \pi_{\mbox B}}}}

\newcommand {\JpinsApibs}{J^{{\tiny \pi_{\mbox A}^{\mbox{\tiny NS}}, \pi_{\mbox B}^*}}}

\def\bsym#1{{\boldsymbol{#1}}}

\newtheorem{assum}{Assumption}

\begin{document}

\title{\vspace{20mm}\sffamily{\bfseries Play Like the Pros? Solving the Game of Darts as a Dynamic Zero-Sum Game} } 

\author{\sffamily Martin B. Haugh \\
\sffamily Department of Analytics, Marketing \& Operations \\
\sffamily Imperial College Business School, Imperial College \\
\texttt{m.haugh@imperial.ac.uk}
\vspace{.75cm}
\and \sffamily Chun Wang \\
\sffamily  Department of Management Science and Engineering\\
\sffamily School of Economics and Management, Tsinghua University \\
\texttt{wangchun@sem.tsinghua.edu.cn}}

\date{\today}

\maketitle

\centerline {\large \bf Abstract} \baselineskip 14pt
\noindent
The game of darts has enjoyed great growth over the past decade with the perception of darts moving from that of a pub game to a game that is regularly scheduled on prime-time television in many countries including the U.K., Germany, the Netherlands, and Australia among others. The game of darts involves strategic interactions between two players but to date the literature has ignored these interactions. In this paper, we formulate and solve the game of darts as a dynamic zero-sum-game (ZSG), and to the best of our knowledge we are the first to do so. We also estimate individual skill models using a novel data-set based on darts matches that were played by the top 16 professional players in the world during the 2019 season. Using the fitted skill models and our ZSG problem formulation, we quantify the importance of playing strategically, i.e. taking into account the score and strategy of one's opponent, when computing an optimal strategy. For top professionals we find that playing strategically results in an increase in win-probability of just 0.2\% - 0.6\% over a single leg but as much as 2.3\% over a best-of-35 legs match.

\clearpage
\onehalfspacing

\section{Introduction}
\label{sec:intro}

In recent years the game of darts has experienced a surge in popularity with the game now having a substantial presence in many countries including the U.K., Ireland, Germany, the Netherlands, and Australia among others. Indeed a recent headline in \cite{Economist2020} explains ``How darts flew from pastime to prime time'' and how it is not uncommon today to have tens of thousands of fans attend darts tournaments. The game is also becoming increasingly attractive to women as evidenced by the exploits of Fallon Sherrock who in 2019 became the first woman to win a match at the PDC World Darts Championship and who made it to the quarter-final of the Grand Slam of Darts in 2021. The aforementioned article from \texttt{The Economist} also notes how in recent years darts has become the second-most-watched sport over the Christmas period on Sky Sports in the U.K., coming second only to soccer. (Sky Sports is the premium sports television channel in the U.K. and is comparable to ESPN in the U.S.)  Further growth is also anticipated with professional tournaments now taking place in Shanghai and scheduled to take place in Madison Square Garden in New York in 2022.

At its core, darts is a game between two players who both start on a fixed score (typically 501) and take it in turns to throw three darts at a dartboard approximately 8 feet away. Loosely speaking, each dart scores a number of points depending on where it lands on the board and these points are then subtracted from the player's score to yield an updated score. The first player to have an updated score of exactly zero wins the game or {\em leg} but the winning dart must be a so-called double.
Darts is therefore a dynamic zero-sum game (ZSG) game where two players race each other to a score of zero.  This leads to two approaches to modeling the game of darts:
\be
\item Formulate a Markov Decision Process (MDP) problem where the goal of a player is simply to minimize the expected number of turns it takes to reach zero. This is sub-optimal since the optimal policy for the MDP takes no account of the score of his/her (hereafter ``his'') opponent. Indeed it is well-known that top players do take account of their opponent's score when they are playing.

\item Formulate the game of darts as a dynamic ZSG and solve for the equilibrium strategies of both players. Alternatively, if we believe our opponent will play some fixed policy, then we would like to solve for the {\em best-response} (BR) policy against his fixed policy.
\ee
Because the game of darts involves strategic interactions between the two players, it should be modeled as a dynamic ZSG, but to date the literature has ignored these interactions and only considered MDP formulations. The central contribution of our work is therefore to formulate and solve the game as a ZSG.

In order to distinguish between the MDP and ZSG approaches, we will use the term ``strategic'' when referring to the BR/ZSG policies, and will generally use the term ``non-strategic'' (NS) when referring to optimal MDP policies that do not account in any way for the opponent's score or strategy. (This is a slight abuse of notation as MDP policies are strategic in the sense that they do take account of a player's own score.)

Solving for BR and equilibrium policies is considerably more challenging than simply taking the first approach and modeling the problem as an MDP. Therefore, one of the central goals of this paper is to formulate and solve the BR and ZSG problems for a darts {\em leg}, i.e. a single race to zero. A further goal is to use the resulting ZSG solution to quantify the extent of sub-optimality in following the MDP/NS approach.
We suspect the extent of this sub-optimality to be quite small on any given leg, since it is only when the players get relatively close to zero that it may be necessary to account for an opponent's score when determining the optimal throwing policy. Nonetheless, no one to date has quantified the extent of this sub-optimality. Moreover, it is possible that a small sub-optimality gap on a single leg can translate to a more considerable gap across a full darts match which typically consists of many legs.

In order to tackle these goals we borrow and extend the skill modeling approach of \cite{Tibshirani} (hereafter TPT). We use a novel data-set based on the dart-throws of the top 16 professional players from the 2019 season to fit a simple extension of TPT's bivariate normal skill model. With estimated skill models for each of the 16 players, we then formulate and solve the game of darts as a dynamic ZSG. Using the fitted skill models, we construct NS, BR, and ZSG equilibrium strategies for each of the players.
We quantify the difference in win-probability between playing the ZSG equilibrium strategy instead of the NS strategy against a player playing the ZSG equilibrium strategy.
Over a single leg this difference is quite small and on the order of approximately 0.2\% - 0.6\% for the top professional players.
However, a darts match is typically played across many legs, and then this 0.2\% - 0.6\% difference across a single leg can translate to a difference of as much as 2.3\% across a best-of-35 legs match which is typical in many big tournaments. We are therefore the first to quantify the strategic element, i.e. the importance of taking an opponent's score into account, when playing the games of darts.

To date, there has been relatively little analytic work on the game of darts in the literature although this has begun to change in recent years. The earliest paper that we are aware of is the work by \cite{Kohler} who uses an MDP formulation and a branch-and-bound approach to solve a slightly simplified version of the game. More recently \citet{Baird2020} considered an MDP formulation where the goal is to minimize the expected number of turns until {\em checking out}, i.e. reaching a score of zero with the last throw being a double. Both \cite{Kohler} and \citet{Baird2020} therefore ignore their opponent's score when constructing a policy and produce sub-optimal policies for the ZSG.

\cite{Stern-CHANCE} and \cite{Percy99} considered the easier problem of where on the dartboard
to aim in order to maximize the score of an individual dart. The former used the Weibull distribution to model and fit dart throws. More recently TPT proposed several models based on the Gaussian and skew Gaussian bivariate distributions to model the throwing skills of players. They use the EM
algorithm together with importance sampling to fit these distributions to dart-throwing data.

Another recent development is the work by \citet{HotHands_RSSA} who find some evidence for a weak-form of the so-called ``hot-hand'' phenomenon in darts. In particular, they find strong evidence for it during the three dart throws {\em within} a turn, but they find little evidence for it persisting {\em across} turns. (In Section \ref{sec:Rules} we clarify what we mean by a ``turn''.)

More generally the OR/MS community has taken a great interest in data analytics over the past couple of decades and concurrent with this has been a great interest in sports analytics; see for example \citet{Wright2009} and more recently \citet{ORMSToday2017}. Some specific examples of work in the field include  \cite{KaplanMarchMadness} on office pools motivated by the ``March madness'' NCAA basketball tournament, \cite{KaplanHockey} on ice hockey, \cite{ChanSingal2016} on handicapping tennis, \cite{Hendershott} on assigning possession of the football in the NFL, and more recently  \cite{hunter2019picking}, \cite{BergmanOR2017} and \cite{HaughSingal} on fantasy sports.  We see this paper as adding to this growing literature on OR/MS applications in sports.

Finally, this paper also relates to the substantial literature on dynamic ZSGs. ZSGs occur in many contexts including interdiction games, e.g. \citet{InterdictionGames2019} and \citet{AirDefense2016}, the management of communication networks, e.g. \citet{DiverseRouting2002}, and the modeling of heads-up poker which is one of the most important test arenas for the development of artificial intelligence (AI) research. Great progress has been made in recent years and in fact AI-based poker players are now capable of beating the best professionals in the world; see for example \citet{Bowling145} and \citet{Sandholm122}.

The remainder of this paper is organized as follows. In Section \ref{sec:Rules} we explain the rules of darts while in Section \ref{sec:SkillModel} we describe our conditional Gaussian skill-model for throwing darts.
In Section \ref{sec:ZSG-formulation} we formulate the ZSG problem and then in Section \ref{sec:Solve} we describe our algorithm for solving it.
We present our numerical results in Section \ref{sec:Numerics} and we conclude in Section \ref{sec:Conclusions} where we also outline some directions for future research. Various technical details as well as additional numerical results are provided in the appendices and online appendices.

\section{The Rules of Darts}
\label{sec:Rules}

\bigskip
\noindent
\begin{minipage}[b]{0.55\linewidth}
In order to understand the game of darts, we need to understand how a single dart scores.
Figure \ref{fig:DartB1} displays a standard dartboard to aid our discussion.
The score obtained by a dart depends entirely on where on the board the dart lands. 
In particular, the dart-thrower does not declare a target although he will invariably have an (undeclared) {\em intended} target. The small concentric circles in the middle of the board define the ``double bulls-eye'' (DB) and the ``single bulls-eye'' (SB) regions. If a dart lands in the DB region (the small red circle) then it scores 50 points while a dart landing in the SB region (the green annulus surrounding the DB region) scores 25.
\end{minipage} \hspace{.5cm}
\begin{minipage}[t]{0.45\linewidth}    \vspace{-6.75cm}
\centering
\includegraphics[width=.8\linewidth]{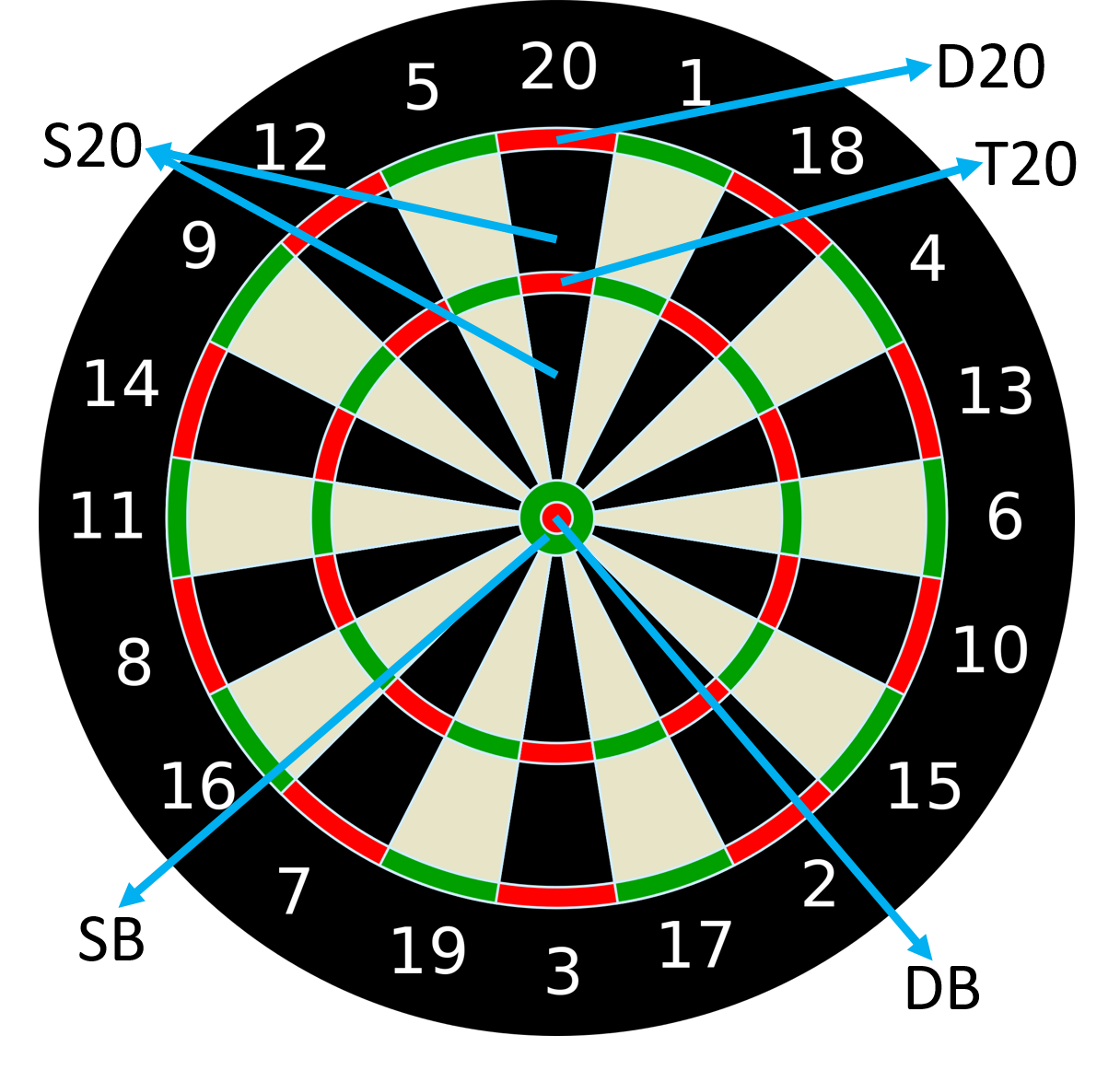}
\captionof{figure}{A standard dartboard}
\label{fig:DartB1}
\end{minipage}
\noindent

Beyond the SB region, the dartboard is divided into 20 segments of equal area corresponding to the numbers 1 to 20. If a dart lands in the ``20'' segment, for example, then it will land in the treble twenty (T20) region, the double twenty region (D20), or the single twenty region (S20) for scores of 60, 40, or 20, respectively. The double region is the region between the two outermost circles on the dartboard whilst the treble region is the region between the two circles beyond the SB region.
The single region is then the union of the two disjoint regions between the SB and treble regions and between the treble and double regions.
If a dart lands beyond the double region or simply fails to land on the board then it scores zero. (The distances from the center to the various regions on the dartboard are provided in Appendix \ref{sec:geom}.)

The rules of darts that we describe here
are for the most commonly played form of the game, namely ``501''. This is the form of the game typically played in pubs and most professional tournaments. In 501 each player starts on a score of 501 and takes turns in throwing three darts. These three darts constitute a ``turn''. After a player's turn, his scores on the three darts are added together and subtracted from his score at the beginning of the turn. The first player to reach a score of {\em exactly} zero wins the game as long as his final dart is a double, i.e. a dart that lands on D1, D2, ..., D20, or DB in which case we say the player has {\em checked out}. If a player's turn would result in an updated score of either 1, 0 (but not achieved via a double on his last dart) or a negative score, then we say the player has gone {\em bust}. In this case the turn is invalidated, the player remains on the score he had prior to the turn and his opponent then takes his turn.  Note that it is possible for a player to win without having to throw all three darts in his turn. For example, suppose a player has a score of 20 just prior to his turn. If he then scores D10 with the first dart of his turn he wins. Alternatively, if he missed D10 with his first dart and scored S10, then he could still win with the second dart of his turn by throwing a D5.

The game we have just described is known as a leg. In practice, darts matches are typically played over many legs with the winner being the first to win some fixed number of legs. Alternatively, some tournaments have a legs and ``sets'' structure whereby the winner of the match is the first to win a fixed number of sets and the winner of a set is the first to win a fixed number of legs. This latter structure then is similar to other sports such as tennis.  Finally, because the player that throws first has an advantage, players alternate in starting legs (and sets). For example, if player A begins the first leg of the first set, then player B will begin the first leg of the second set, etc. Similarly within a set, players alternate in starting each leg.
Regardless of the match structure, however, the key component of a darts match is the leg and it should be clear that a player will maximize his chances of winning a match if he optimizes his strategy for playing each leg.

\section{A Skill Model for Dart Throwing}
\label{sec:SkillModel}

Taking the center of the DB region to be the origin, we will use $\ba \in \mathbb{R}^2$ to denote
the target of the dart throw and $(x,y) \in \mathbb{R}^2$ to denote the outcome of the throw. We then let $p(x,y;\, \ba)$ denote the probability of the outcome $(x,y)$ given the target $\ba$. Given $p(x,y;\, \ba)$ it is then easy to compute the distribution $p(z;\, \ba)$ where $z=g(x,y)$ maps $(x,y)$ to the dart score $z$, e.g. D16, SB, T20, S7, etc. We do this by numerically integrating $p(x,y;\, \ba)$ over the region of the dartboard defined by $z$. For example, to compute $p(\mbox{T20};\, \ba)$ we integrate $p(x,y;\, \ba)$ over the T20 region of the dartboard.
We will also let $h(z) \in \mathbb{N}$ denote the actual numerical score achieved by the dart. For example, $h(\text{D16}) = 32$, $h(\text{T20})=60$, $h(\text{SB}) = 25$, $h(\text{S7}) = 7$, etc. Given $p(z;\, \ba)$ it is also easy to compute the distribution $p(h(z);\, \ba)$. Our goal then is to construct a model for $p(x,y;\, \ba)$ that can also be easily estimated with available data.

TPT proposed several models for capturing the skill of a darts player. Their first model assumed  the outcome $(x,y)$ of the throw follows
a bivariate Gaussian distribution with mean $\ba$ (the intended target) and covariance matrix $\sigma^2 \bI$ where $\bI$ is the $2 \times 2$ identity matrix.
They also proposed the more general model
\begin{eqnarray}  \label{eq:DM2}
(x,y) & \sim & \mbox{N}_2(\ba, \bSigma)
\end{eqnarray}
where $\bSigma \in \mathbb{R}^{2,2}$ is an arbitrary covariance matrix. TPT proposed the use of the EM algorithm to estimate $\bSigma$ in (\ref{eq:DM2}). They assumed the intended target $\ba$ was known for each data-point but that only the result $z=g(x,y)$ was observed rather than the realized location $(x,y)$. That only the $z$'s were observed is a very reasonable assumption given the difficulty of measuring the $(x,y)$-coordinates of dart throws. Indeed this is certainly true for the data-set we will use in our numerical study in Section \ref{sec:Numerics}.

A simple model such as (\ref{eq:DM2}) will not be sufficiently rich for modeling the skills of professional darts players, because these players tend to focus on (and practice throwing at) specific parts of the darts board, e.g. T20, T19, DB, etc. This means that their skill levels, as determined by $\bSigma$, are likely to depend on $\ba$, which is indeed what we observe in the data-set that we use in Section \ref{sec:Numerics}.
We therefore assume there exists a finite number of regions of the board, say $R_1, \ldots , R_M$, so that (\ref{eq:DM2}) can then be generalized to
\begin{eqnarray}
 \label{eq:DM4}
(x,y) & \sim & \mbox{N}_2(\ba, \bSigma_m), \ \ \mbox{ for } \ba \in R_m.
\end{eqnarray}
Note that (\ref{eq:DM4}) is as easy to estimate as (\ref{eq:DM2}) since we can partition our data-set by regions $R_m$ with each region consisting of one or more target regions.

Once we have fitted a model $p(x,y; \ba)$ such as (\ref{eq:DM4}) we can then easily compute $p(z; \ba)$, the corresponding distribution over scores, and $p(h(z); \ba)$, the corresponding distribution over numerical scores.

\section{The Zero-Sum Game Formulation}
\label{sec:ZSG-formulation}

Ultimately a darts match (or specifically a leg of a darts match) is a dynamic ZSG where the two players are in a race to zero. This means that, at least in principle, a player needs to account for his opponent's strategy when choosing his own strategy. For example, it is well known that a player's decision to target the DB will depend in part on his opponent's score.

Before proceeding, we clarify the zero-sum nature of a darts leg. In practice, the winner of the leg scores one point and the loser scores zero and clearly these two numbers do not sum to zero. However, we could scale both prizes by a factor of two and then subtract one from each of them to obtain prizes of $(1,-1)$ which are now in zero-sum format. It should be clear that this transformation alters nothing in the players' objectives and so a leg of darts can indeed be modelled as a zero-sum game. We will express the game objective from the perspective of player A who wishes to maximize the probability that he wins, i.e. that he reaches zero before player B. Player B therefore wishes to minimize this probability.

These win-probabilities will be a function of the state $\bs = (\sA,\sB,t,i,u) \in {\cal S}_{\mbox{\tiny ZSG}}$, 
where $\sA$ and $\sB$ are the scores at the beginning of the turn for A and B, respectively,
$t \in \{\mbox{A},\mbox{B}\}$ denotes whose turn it is, $i \in \{1,2,3\}$ denotes how many throws are left in player $t$'s turn, and $u$ is player $t$'s cumulative score thus far within the current turn.
The set ${\cal S}_{\mbox{\tiny ZSG}}$ then denotes the state-space for the ZSG. Because we are taking player A's perspective, the rewards for the ZSG satisfy $r(\sA,\sB,t,i,u) = 1_{\{\sA =0 \}}$, where $1_{\{\}}$ is the indicator function.
The states $\{\sA =0 \}$ and $\{\sB =0 \}$ correspond to A winning and B winning, respectively.
Technically, we do not want to have $\{\sA =0 \}$ as an absorbing state as this would imply A receives a reward of 1 for every period in perpetuity. Instead we assume that after reaching the states $\{\sA =0 \}$ or $\{\sB =0 \}$ we then immediately move to an absorbing state $\tau$ that is rewardless.

We define our action-space ${\cal A}$ to be the set of possible targets on the dartboard. In our setting, ${\cal A}$ is state-independent and we assume it is finite, reflecting the fact that any human will only be able to reasonably distinguish a finite number of targets on the dartboard. An admissible policy $\pi := \{\pi_1, \pi_2, \ldots \}$ for player $j \in \{\mbox{A},\mbox{B}\}$ is a policy where each $\pi_k$ is a decision rule defining the target of the $k^{th}$ dart. That is, $\pi_k(\bs) \in {\cal A}$ for any $\bs \in {\cal S}_{\mbox{\tiny ZSG}}$ with $t=j$ and $k \in \mathbb{N}^+$ denoting the $k^{th}$ dart thrown in the leg. Recall that $t$ is the component of $\bs$ denoting whose turn it is. If $t \not = j$, then it is the turn of $j$'s opponent and we set $\pi_k(\bs) := \emptyset$ to reflect the fact that he does not throw a dart in this state. Stationary strategies $\pi$ satisfy $\pi_1=\pi_2= \cdots$ and in this case we will use $\pi$ to denote both the (stationary) strategy as well as the decision rule $\pi(\bs)$. We let $\Pi$ denote the space of all admissible policies $\pi$.

Let $\JpiApib (\bs)$ be the probability that player A wins the leg when A and B play strategies $\piA$ and $\piB$, respectively, and where $\bs = (\sA,\sB,t,i,u)$ denotes the current state of the game.
That is, we define
$\JpiApib (\bs):= \lim_{K\to \infty} \Ex \left[ \sum_{k=1}^K r(\bs_k) \mid \bs_1 = \bs\right]$,
where $\Ex$ indicates expectation and $K$ denotes the total number of permitted dart throws in a finite-horizon version of the problem.
The min-max and max-min values of the game are then defined to be
\begin{eqnarray*}
\underline{J}(\bs) &:=& \min_{\piB \in \Pi} \max_{\piA \in \Pi} \JpiApib (\bs) \\
\overline{J}(\bs) &:=& \max_{\piA \in \Pi} \min_{\piB \in \Pi}  \JpiApib (\bs).
\end{eqnarray*}
The study of dynamic ZSG's was initiated in a classic paper by \cite{Shapley1953}. For infinite horizon discounted problems he showed $\underline{J}(\bs) = \overline{J}(\bs) =: J^*(\bs)$ and that there exist stationary strategies $\piA^*$ and $\piB^*$  satisfying
\begin{equation}\label{eq:EquilZSG1}
\JpisApibs(\bs) = \max_{\piA  \in \Pi} \JpiApibs (\bs) =  \min_{\piB  \in \Pi} \JpisApib (\bs) = \underline{J}(\bs) = \overline{J}(\bs) = J^*(\bs).
\end{equation}
It follows from (\ref{eq:EquilZSG1}) that $\piA^*$ and $\piB^*$ are optimal or equilibrium strategies for A and B, respectively,
and we say $J^*(\bs)$ is the equilibrium value of the game starting from state $\bs$.

In our setting, the ZSG is an undiscounted stochastic shortest-path (SSP) ZSG  where the absorbing state $\tau$ (that immediately follows player A or B winning) is rewardless and reached with probability one under all
strategies for A and B, i.e. we assume every strategy is {\em proper}. It is easy to show that this assumption holds in our darts set-up. Under these conditions \cite{Patek97} showed that (\ref{eq:EquilZSG1}) continues to hold and that $J^*(\bs)$ is the unique equilibrium of the ZSG.

\cite{Patek97} actually considered {\em simultaneous}-move ZSGs where the optimal equilibrium strategies are randomized. Our darts ZSG is an {\em alternate}-move ZSG but it is easy to write this as a simultaneous-move game. Specifically, we can imagine introducing a {\em dummy} action for each player and assuming the dummy action is the only action available to a player when it is in fact his opponent's turn to throw. (Essentially, this is what we did above when we set $\pi_k(\bs) := \emptyset$ for player $j$ when the $t$ component of $\bs$ is not equal to $j$.) This dummy action has no effect on the state of the game. Moreover, this structure where in every stage one of the player's feasible action sets is a singleton (dummy) action results in equilibrium strategies that are non-randomized.

\section{Solving the ZSG}
\label{sec:Solve}

We now describe our approach to solving the ZSG (\ref{eq:EquilZSG1}). As noted above, \cite{Patek97} showed that (\ref{eq:EquilZSG1}) holds in our ZSG setting and that $J^*(\bs)$ is the unique equilibrium of the game. Moreover, they also showed that value iteration will converge to the unique optimal solution, and that is the approach we take in this paper.
Specifically, we will initialize our value iteration by first computing player B's optimal NS strategy $\piNSB$. We then compute A's best-response strategy to $\piNSB$ and we call this $\piBRA$. We can then compute B's best-response to $\piBRA$ and iterate in this manner until convergence.

Moreover, convergence is easy to assess since solving player A's (B's) best-response to any player B (A) strategy leads to an upper (lower) bound on the value of the game $J^*(\bs)$, i.e. the win-probability for player A when both players play their equilibrium strategies. This follows immediately from (\ref{eq:EquilZSG1}). At each value iteration, we can therefore compute lower and upper bounds on the optimal value $J^*(\bs)$ and stop when these bounds are sufficiently close to each other. In our darts setting, we find that this typically takes only two or three iterations.

In Section \ref{sec:Solve-BRs} we formulate and solve the best-response problem, and then in Section \ref{sec:Solve-ZSG} we describe our value iteration algorithm for solving the ZSG in more detail. Appendix \ref{sec:DP-formulation} contains the NS model formulation as well as our algorithm for computing the NS strategy.

\subsection{The Best-Response Problem}
\label{sec:Solve-BRs}

We now solve for player A's best-response strategy to player B's strategy $\piB \in \Pi$. We therefore assume $\piB $ is fixed in this subsection. Recall that the state of the game at any time is denoted by $\bs = (\sA,\sB,t,i,u)$. However, we can avoid including $t \in \{\mbox{A},\mbox{B}\}$ as a component of $\bs$ by assuming it is always player A's turn and noting that player B's score will change (randomly according to B's strategy $\piB$) at the end of A's turn. With this new definition of ${\cal S}_{\mbox{\tiny ZSG}}$, we define the state transition function for A's best-response as
\begin{equation}  \label{eq:Dynamics-DP-ABR}
 f((\sA,\sB,i,u),z,\snewB) := \left\{
            \begin{array}{ll}
              (0,\sB,0,0), &  \sA = u + h(z) \mbox{ and } z \mbox{ a double, } i\in\{1,2,3\}\\
              (\sA,\snewB,3,0), & \sA - (u + h(z)) \leq 1 \mbox{ and not }\\
              & \hfill (\sA=u+h(z), \ z \mbox{ a double}),   i\in\{1,2,3\}\\
              (\sA,\sB,i-1,u+h(z)), & \sA > (u + h(z)) +1, \ i > 1 \\
              (\sA-u-h(z),\snewB,3,0), & \sA > (u + h(z)) + 1, \ i = 1
            \end{array}
          \right.
\end{equation}
\noindent
where $h(z) \geq 0$ is the random score of the dart. Player A wins the leg in case 1 of (\ref{eq:Dynamics-DP-ABR}), while in case 3 he still has at least one more throw in his turn and has not gone bust. Player B's updated score after A's turn ends is denoted by $\snewB$, and it must satisfy $ 0 \leq \snewB \leq \sB$. So $\snewB$ is just the result of B's turn which takes place between successive turns of A. We observe that B's score updates to $\snewB$ only in cases 2 and 4 of (\ref{eq:Dynamics-DP-ABR}) when A has either gone bust (case 2) or ended his turn without winning (case 4).

Player A wins the leg and therefore obtains a reward of 1 only in the state $(0,\sB,0,0)$, i.e. in case 1 of (\ref{eq:Dynamics-DP-ABR}), after which we assume the state immediately moves to the rewardless absorbing state $\tau$. The dynamics in (\ref{eq:Dynamics-DP-ABR}) implicitly assume that $\snewB \not = 0$. However, it should be understood that any transition where $\snewB=0$ implies that B wins the leg and is immediately followed by a transition to $\tau$.

Let $\VA(\bs)$ denote player A's best-response value function to player B's strategy $\piB \in \Pi$ for arbitrary states $\bs = (\sA,\sB,i,u) \in {\cal S}_{\mbox{\tiny ZSG}}$.
Then, player A's best-response can be formulated as the MDP
\begin{equation} \label{eq:BR-PRobForm1}
\VA(\bs_1) = \max_{\piA \in \Pi} \JpiApib (\bs_1)
\end{equation}
where $\bs_1 := (501,501,3,0)$ is the initial state of the game. For ease of exposition we are assuming A is first to throw in the leg so B starts on a score of 501 after A's first turn. We have also suppressed the dependence of $\VA$ on $\piB$ to ease the notational burden.

Problem (\ref{eq:BR-PRobForm1}) is an SSP and standard dynamic programming (DP) techniques can be used to show that it has a unique solution.
For $\sA,\sB > 0$, it satisfies the Bellman equation
\begin{equation}\label{eq:DP-ABR1}
\VA(\sA,\sB,i,u) = \max_{\ba \in {\cal A}} \left\{ \Ex\left[\VA( f((\sA,\sB,i,u),z,\snewB))  \mid (\sA,\sB,i,u),\ba \right] \right\}
\end{equation}
with $\VA(0,\sB,0,0)=1$ (A winning) and $\VA(\sA,0,3,0) =0$ (B winning) and where the expectation in (\ref{eq:DP-ABR1}) is taken with respect to $z$ and $\snewB$.
Value iteration converges linearly for this problem. In particular, for a given error tolerance  $\epsilon$ under a suitably weighted norm (see e.g. \citealp{Bert-Yu-2016}), we can solve (\ref{eq:DP-ABR1}) in $O(|{\cal S}_{\mbox{\tiny ZSG}}| \times |{\cal A}| \times \log (1/\epsilon))$ time. (We require $O(\log (1/\epsilon))$ value iterations and each iteration requires us to find the optimal action in each of $|{\cal S}_{\mbox{\tiny ZSG}}|$ states.  As there are $|{\cal A}|$ possible actions to consider for each state, each iteration therefore requires $O(|{\cal S}_{\mbox{\tiny ZSG}}| \times |{\cal A}|)$ time.) We need $O(\max \{|{\cal S}_{\mbox{\tiny ZSG}}|,\,  |{\cal A}|\})$ space for solving these problems.

There is a monotonic structure to the DP (\ref{eq:DP-ABR1}): specifically, it is clear from the rules of the game, and indeed from the state transitions in (\ref{eq:Dynamics-DP-ABR}), that the $\sA$ and $\sB$ components of the state $(\sA,\sB,i,u)$ are non-increasing. We can use this monotonic structure to solve (\ref{eq:DP-ABR1}) more efficiently. In particular, we proceed via Algorithm \ref{alg:BR}:
\begin{algorithm}[H]
\caption{Pseudo-code for solving the best-response DP}
\label{alg:BR}
\begin{algorithmic}
\FOR{$\sB = 2:501$}
\FOR{$\sA = 2:501$}
\STATE{Solve for $\VA(\sA,\sB,i,u)$ via (\ref{eq:DP-ABR1}) for all $(i, u)$}
\ENDFOR
\ENDFOR
\end{algorithmic}
\end{algorithm}\noindent
We used policy iteration to solve the DPs inside the inner for-loop of Algorithm \ref{alg:BR}. For each such DP, we used the previously computed optimal policies (from earlier iterations of the for-loops) to quickly find the optimal policy. We also experimented with using value iteration for solving these BR problems but found it to be considerably slower than policy iteration.
(While our running-time complexity was given for value iteration, we found policy iteration to be faster. Policy iteration is also guaranteed to converge but we are not aware of a running-time complexity result for it that we can use for our SSPs.)

\subsection{Solving the Zero-Sum Game}
\label{sec:Solve-ZSG}

Now that we can solve for best-response strategies, we turn our attention to the value iteration algorithm for solving the ZSG, i.e. solving for $J^*(\bs)$ as defined in (\ref{eq:EquilZSG1}).
As mentioned at the beginning of the section, value iteration begins by fixing an initial strategy for one player, e.g. $\piNSB$ for player B, and finding A's best response $\piBRA$ to this strategy. This best-response strategy can be computed by solving (\ref{eq:DP-ABR1}) via Algorithm \ref{alg:BR}. We then fix A's strategy $\piBRA$ and find B's best-response to $\piBRA$.
We continue to iterate, alternately finding A's best-response to B and B's best-response to A, until convergence.
The results of \cite{Patek97} imply that value iteration for the ZSG converges linearly; cf. the paragraph immediately following Prop. 4.6 in \cite{Patek97}.
Linear convergence implies $O(\log (1/\epsilon))$ such iterations will be required to solve for the optimal ZSG value functions within an error tolerance of $\epsilon $ (again under a suitably weighted norm).

As with the best-response DP formulation of Section \ref{sec:Solve-BRs}, we can again take advantage of the monotonicity structure to accelerate convergence. In particular, we used the following algorithm:
\begin{algorithm}[H]
\caption{Pseudo-code for solving the ZSG exploiting the monotonicity structure}
\label{alg:ZSG}
\begin{algorithmic}
\FOR{$\sB = 2:501$}
\FOR{$\sA = 2:501$}
        \STATE{Set $\piB = \piNSB$, $\bs = (\sA,\sB,\text{A},3,0)$  \hfill \# initialization}
        \WHILE{(not converged)}
            \STATE{Use Alg. \ref{alg:BR} to solve for $\piA$ = A's BR to $\piB$ for states in the turn starting from $\bs$}
            \STATE{Use Alg. \ref{alg:BR} to solve for $\piB$ = B's BR to $\piA$ for states in the turn starting from $\bs$}
            \ENDWHILE
        \STATE{Save $\piA(\bs)$, $\piB(\bs)$, $J^*(\bs)$ }
\ENDFOR
\ENDFOR
\end{algorithmic}
\end{algorithm} \noindent
The policy iteration algorithms used to solve the BR problems inside the while-loop of Algorithm \ref{alg:ZSG} were initialized using the optimal policy from earlier iterations of the for-loops in Algorithm \ref{alg:ZSG}.

\section{Numerical Study}
\label{sec:Numerics}

In this section, we present results obtained by solving our NS and ZSG models using fitted skill models for each of the top 16 professional players in the world during the 2019 season. We fitted these skill models using a novel data-set that we describe below in Section \ref{sec:Data}.
In Section \ref{sec:RunTimes} we describe the software/hardware setup that we used for our numerical investigation, as well as the associated run-times required for executing our algorithms.
In Section \ref{sec:QuantStrat} we use our
solutions to quantify the value for top players of playing strategically in a single leg and over the course of a darts match consisting of several legs.
In Section \ref{sec:Hypo} we use a hypothetical match-play situation involving two of the professional players to demonstrate the kind of analysis for which our BR / ZSG solutions can be used.

\subsection{The Data-Set}
\label{sec:Data}

Our data-set pertains to matches that were played by the top 16 professional players in the world during the 2019 season.
For each of the 16 players we have data of the form (TR, $z$, $n$) where $n$ is the number of darts that were aimed at the {\em target region} (TR) and achieved a score of $z$. There are a total of 62 possible target regions, namely the single regions S1, ..., S20, the double regions D1, ..., D20, the treble regions T1, ..., T20, and the single and double bulls-eye, i.e. SB and DB.
Because many regions of the dartboard are very rarely targeted, the target regions that appear in our data-set are the treble regions T20, T19, T18, and T17 together with all the double regions D1, ..., D20, and the double bulls-eye DB.

The possible realized value of the score $z$ depends on the target region TR since a professional darts player will only very rarely miss his target region TR by more than a small distance.
For each treble region there are therefore 6 possible $z$ scores. In the case of TR = T20, for example, the possible values are $z \in \{\text{T20, S20, T5, S5, T1, S1} \}$ because the 5 and 1 segments of the dartboard are adjacent to the 20 segment; see Figure \ref{fig:DartB1}. Because we only have data for 4 treble regions, this means a total of $6 \times 4 = 24$ data-vectors (TR, $z$, $n$) for each player where TR is a treble region.
For each double target region TR = D$x$ for $1 \leq x \leq 20$, we have 7 possible $z$ values represented in the data-set. For example, if TR = D16 then the corresponding possible values of $z$ are $z\in \{\text{D16, S16, D8, S8, D7, S7, M}\}$ because 8 and 7 are adjacent to 16 on the dartboard and where M denotes a ``miss'', i.e. a dart that fails to score because it landed outside the double region; again see Figure \ref{fig:DartB1}.
When TR = DB we have 22 possible $z$'s corresponding to \{DB, SB, S1, S2, ..., S20\}.
Because targeting the DB and some of the double regions is relatively rare, we note that for some of the (TR, $z$, $n$) combinations we occasionally have $n=0$.

Unfortunately, when a player targets a particular region, e.g. T20, we do not know the exact point in the region at which he was aiming. We therefore make the following assumption.
\begin{assum} \label{ass:Center}
Given a target region TR, the specific target that a player aims for is the center of the target region with the center being defined as the midpoint of the polar-coordinates defining the region.
\end{assum}
We therefore assume the precise target location
for each target region TR is known to us. While this assumption may not always be true, absent any other information it seems like the most natural assumption to make. Indeed it is not difficult to imagine a player deviating from this assumption on occasion. For example, suppose a player has two darts remaining in his turn and needs a D5 to check out, i.e. win the game. Rather than aiming at the center of D5, he may prefer to aim a little closer to the outer circular boundary of the D5 region on the basis that if he is going to miss D5 he would prefer to miss the scoring region entirely rather than hit S5 which would leave him unable to check out on his final dart of the turn. This argument does not apply to even doubles such as D20, D18, D16, etc. since if the corresponding single is hit then the player can still exit on the next dart in his turn.
We also make two further assumptions regarding so-called ``bounce-outs'' and the players' within-turn skill models.
\vspace{-.1cm}
\begin{assum} \label{ass:BounceOut}
There is no possibility of a bounce-out occurring when a dart is thrown.
\end{assum}

A ``bounce-out'' is a failure of the dart to actually land in the dartboard. This can occur because of a poor throw or because the tip of the dart strikes the wire that defines the boundaries between the different regions. With a sufficiently sharp dart tip, however, this rarely happens and in fact only approximately 3 in every 1,000 darts thrown by top players were bounce-outs in the 2019 season.

\begin{assum} \label{ass:FORM}
Each player has the same skill model for each throw in his turn.
\end{assum}
It is known (see, for example, \citealp{HotHands_RSSA}) that the success rate of the first dart in a turn is generally smaller than the success rate of the second and third throws of the turn. This can be explained by the need of the dart-thrower to ``re-calibrate'' at the beginning of each turn. Unfortunately, our data-set is not sufficiently granular to estimate a different skill model for each throw in a turn and so we need to make Assumption \ref{ass:FORM}.
Of course if we had sufficiently granular data then for each player we could fit one skill model for the first throw in a turn and a separate skill model for the second and third throws. These skill models could then easily be used in our models with almost no increase in computational requirements.

Our data-set for a given player therefore takes the form $\{(\ba_i,z_i,n_i )\}_{i=1}^p$ where $p$ is the number of unique (TR, $z$) combinations with $n >0$ for that player. We follow TPT and use the EM algorithm to estimate each component of the skill model in (\ref{eq:DM4}) for each of the sixteen players. The details of the EM algorithm are provided in Online Appendix \ref{app:EM}.

In this paper, we assumed the skill model (\ref{eq:DM4}) with $M=6$. These six regions are $R_1 = \{\mbox{T}20\}$, $R_2 = \{\mbox{T}19\}$, $R_3=\{\mbox{T}18\}$, $R_4 = \{\mbox{T}17\}$, $R_5 = \{\mbox{DB}\}$ and $R_6 = \{\mbox{D}1, \ldots , \mbox{D}20 \}$.
Only $R_6$ then consists of more than one target region. This choice was largely governed by the available data. For example, while we have a lot of data on the high trebles, i.e. T20 to T17, we have relatively less data on the doubles. To be clear, we do have a reasonable amount of data on some particular doubles, e.g. D20 and D16, but we have relatively little data on many of the others. Rather than fitting individual skill models to D20 and D16 as we did with T20, T19, T18, and T17, we instead decided to fit a single skill model to all of the doubles.

\subsection{Software/Hardware, Run-Times and Memory Requirements}
\label{sec:RunTimes}

All of our numerical results were obtained using Python running on an Intel Xeon E5-2620 v4 machine with a 2.1GHZ CPU and an Nvidia GeForce GTX 1080 Ti GPU. The size of the state-space ${\cal S}_{\mbox{\tiny ZSG}}$ for solving a best-response problem was on the order of $501^2 \times (1+61+121) \approx 45$ million. ($1$, $61$ and $121$ are the number of possible $u$ values when the number of throws $i$ remaining in a turn is $3$, $2$ and $1$, respectively. Therefore $1+61+121$ is the number of $(i,u)$ combinations possible in a given state.) We also needed to specify the space ${\cal A}$ of permissable actions, i.e. targets $\ba$. We adopted a brute force approach and divided the dartboard into a grid of 1mm $\times$ 1mm squares and took ${\cal A}$ to be the centers of each of these squares. Since the dartboard has a radius of 170mm, this resulted in an action-space with approximately $3.14 \times 170^2 = 90,785$ possible targets. (We also used this 1mm $\times$ 1mm discretization for the numerical integrations required to compute $p(z;\, \ba)$ from $p(x,y;\, \ba)$ as discussed in Section \ref{sec:SkillModel}.)
When computing  $J^*(\bs)$ for a ZSG with this action set, Algorithm \ref{alg:ZSG} converged to a relative error of $10^{-9}$ after just 2 or 3 value iterations for a majority of the states.
Each value iteration step, however, needs to examine every possible target in ${\cal A}$ to determine a player's best target and this must be done for each state $\bs$. As a result Algorithm \ref{alg:ZSG} originally took hundreds of hours to run when implemented naively. Through the use of vectorization and GPU computation techniques, however, we were able to obtain substantial improvements and ultimately solve a ZSG via Algorithm \ref{alg:ZSG} in approximately 5 hours. Online Appendix \ref{app:SpeedUps} provides further details regarding these improvements.

We might have achieved considerable additional improvements, however, if we took advantage of specific knowledge regarding the game of darts. For example, early in the leg the players only ever consider aiming for T20. Similarly, single regions are only ever considered when a player is relatively close to checking out. These examples and many others could be used to drastically reduce the size of the possible action spaces (by making them state-dependent) and might help us to achieve substantial speedups.

A link to the data and Python code that we used for our numerical study is available at  \url{https://github.com/wangchunsem/OptimalDarts}.

\subsection{Quantifying the Value of Playing Strategically}
\label{sec:QuantStrat}
We compute the leg win-probability for each player playing their NS, equilibrium, or BR (to NS) strategies against every other player over a single leg.
In particular, we consider the six combinations E-E N-N N-E E-N N-B B-N where E, N, and B denote the equilibrium, NS, and BR (to NS) strategies, respectively.
So for example, E-E denotes a ZSG where both players play their equilibrium strategies, N-E denotes a ZSG where player A plays his NS strategy and player B plays his equilibrium strategy, and N-B denotes a ZSG where player A plays his NS strategy and player B plays his BR to player A's NS strategy.
The results are displayed in Tables \ref{table:ProbabilityWinning1} and \ref{table:ProbabilityWinning2} in Online Appendix \ref{sec:WinProbsByStratsLegs}.
The main takeaway from our analysis is that the strategic aspect of the game {\em when played over a single leg} is not substantial. Indeed, across all players the difference in win-probability for player A when N-E is played instead of E-E is on the order of just 0.2\% to 0.6\%. Put another way, if a player  plays his NS strategy (and therefore always ignores the score of his opponent) rather than his equilibrium strategy, then the reduction in his probability of winning the leg varies by just 0.2\% to 0.6\%.

These numbers, however, may misstate the value of playing strategically in the real world where matches
are played over multiple legs (or sets) rather than just a single leg. For example, a match might easily be played as the best of 21 legs or 35 legs and a small edge in win-probability over a single leg may translate into a considerably larger edge in win-probability over a match consisting of multiple legs.
To investigate this possibility,
we will assume for the remainder of this section that player B always plays his equilibrium strategy $\piB^*$.
We know that player A has a small advantage on the order of approx 0.2\% to 0.6\% over a single leg when he plays $\piA^*$, his equilibrium strategy, rather than $\piNSA$, his NS strategy.
But what is the cumulative advantage of A playing $\piA^*$ rather than $\piNSA$ over a full match consisting of $N$ legs where the players alternate in starting each leg? To address this,
let $\PA(\piA,N)$ be the probability that A wins such an N-leg match when A plays $\piA$ throughout the match and A starts the first leg. (We could also consider the case where B starts the first leg but the difference will be negligible for realistic values of $N$.)
We are therefore interested in evaluating A's increase in probability of winning the match when he plays $\piA^*$ rather than $\piNSA$ against $\piB^*$. We define this increase or {\em gain} in win-probability as
\begin{equation} \label{eq:Adv1}
\mbox{Gain}(\piA^*,\piNSA,N) := \PA(\piA^*,N) - \PA(\piNSA,N)
\end{equation}
where we omit the dependence on $\piB^*$ as this is fixed throughout. Each term in (\ref{eq:Adv1}) is easily calculated as discussed in Online Appendix \ref{app:MatchWinProbs}.

\begin{table}[ht]
\small
\centering
\captionsetup{justification=centering}
{
\renewcommand{\tabcolsep}{1mm}
\caption{\small $\mbox{Gain}(\piA^*,\piNSA,N)$ in a Match of $N=31$ Legs}
\label{table:Match_leg31}
\begin{tabular}{l  cccc cccc cccc cccc}
\toprule
~ & p1 & p2 & p3 & p4 & p5 & p6 & p7 & p8 & p9 & p10 & p11 & p12 & p13 & p14 & p15 & p16 \\
\midrule
p1&0.9&0.9&0.9&0.3&1.0&0.9&0.7&0.9&0.7&0.9&0.8&0.9&1.0&0.8&0.6&0.9 \\
p2&1.4&1.4&1.4&0.6&1.4&1.4&1.0&1.3&1.1&1.4&1.3&1.4&1.5&1.3&0.9&1.4 \\
p3&1.1&1.1&1.1&0.4&1.2&1.1&0.9&1.0&0.7&1.2&1.0&1.0&1.1&1.0&0.6&1.1 \\
p4&0.8&0.8&0.6&1.4&0.6&1.1&0.2&1.1&1.5&0.5&0.9&1.1&1.1&1.0&1.5&1.0 \\
p5&1.8&1.8&1.9&0.6&2.0&1.7&1.5&1.6&1.2&1.9&1.6&1.7&1.8&1.6&1.0&1.8 \\
p6&1.5&1.5&1.4&0.8&1.5&1.7&0.9&1.6&1.4&1.4&1.5&1.6&1.7&1.5&1.3&1.6 \\
p7&1.2&1.1&1.3&0.2&1.3&0.9&1.5&0.8&0.5&1.4&0.9&0.9&1.0&0.8&0.4&1.0 \\
p8&1.8&1.8&1.7&1.2&1.8&2.1&1.0&2.0&1.9&1.6&1.8&2.0&2.2&1.9&1.7&2.0 \\
p9&0.6&0.6&0.5&0.6&0.6&0.8&0.3&0.7&0.8&0.5&0.6&0.7&0.8&0.7&0.8&0.7 \\
p10&1.1&1.1&1.2&0.3&1.2&0.9&1.1&0.9&0.6&1.2&1.0&0.9&1.0&0.9&0.5&1.0 \\
p11&1.1&1.1&1.1&0.6&1.1&1.2&0.7&1.1&1.0&1.0&1.1&1.2&1.3&1.1&0.9&1.2 \\
p12&1.3&1.3&1.2&0.8&1.3&1.5&0.8&1.4&1.3&1.2&1.3&1.5&1.6&1.4&1.2&1.5 \\
p13&0.7&0.7&0.6&0.4&0.7&0.7&0.4&0.7&0.6&0.6&0.6&0.7&0.8&0.7&0.6&0.7 \\
p14&1.3&1.3&1.3&0.8&1.3&1.5&0.8&1.4&1.3&1.2&1.3&1.4&1.6&1.4&1.2&1.4 \\
p15&1.1&1.1&0.9&1.4&1.0&1.5&0.4&1.5&1.8&0.8&1.2&1.5&1.6&1.4&1.8&1.4 \\
p16&1.0&1.0&1.0&0.5&1.0&1.1&0.7&1.0&0.9&0.9&0.9&1.0&1.2&1.0&0.8&1.1 \\
\bottomrule
\end{tabular}}~\\
\justify
{\em Note:} Numbers are in percentages. The player on the left column is player A who begins the first leg of the match. The player on the top row is player B who uses his equilibrium strategy $\piB^*$ throughout.
The players' names are provided in Tables \ref{table:ProbabilityWinning1} and \ref{table:ProbabilityWinning2} of Online Appendix \ref{sec:WinProbsByStratsLegs}.
\end{table}

Table \ref{table:Match_leg31} displays $\mbox{Gain}(\piA^*,\piNSA,N)$ for all combinations of players when $N=31$ which corresponds to a typical length match in professional tournaments. We can see that the value of playing strategically ranges from 0.2\% to 2.2\% in this case.
For comparison purposes we also display in Online Appendix  \ref{app:MatchWinProbs} the corresponding tables for 
$N=1$, $21$, and $35$.
For the latter two cases we see that $\mbox{Gain}(\piA^*,\piNSA,N)$ varies from 0.3\% to 1.8\% and from 0.2\% to 2.3\%, respectively. These ranges are comparable with the $N=31$ case displayed in Table \ref{table:Match_leg31} and on average are approximately two to three times higher than they are in the $N=1$ case. While seemingly small, a value of $\mbox{Gain}(\piA^*,\piNSA,N)$ in the 1\%-2\% range is still quite considerable given how closely matched top players are.
This finding suggests that in such matches top players would be at a considerable disadvantage playing an NS strategy rather than an equilibrium strategy against other top players who play their equilibrium strategies.

The function $\mbox{Gain}(\piA^*,\piNSA,N)$ typically monotonically increases in $N$ until some value $N^*$ say, after which it decreases monotonically to zero. This latter behavior is explained by the fact that $\PA(\piA^*,N)$ and $\PA(\piNSA,N)$ will either both converge to 0 or both converge to 1 as $N \to \infty$
{\em unless} playing $\piNSA$ makes A an underdog but switching to playing $\piA^*$ makes him a favorite.
In this latter case we would see $\mbox{Gain}(\piA^*,\piNSA,N)$ monotonically increases to 1 as $N \to \infty$.
Indeed we observed that $\lim_{N \to \infty }\mbox{Gain}(\piA^*,\piNSA,N) \to 1$ in the case of Chisnall (player A) vs. Cross (player B) who, as discussed in Online Appendix \ref{sec:WinProbsByStratsLegs}, are very evenly matched.
This is a case where (at least in the limit of large $N$) taking account of an opponent's score and playing the equilibrium strategy makes all the difference between winning and losing. For opponents as closely matched as this, however, $N$ would need to be very large (and on the order of 1,000's) before we would actually see $\mbox{Gain}(\piA^*,\piNSA,N)$ get very close to 1.

\subsection{Analysis of Matchplay Situations}
\label{sec:Hypo}
\begin{figure}[]
  \begin{center}
    \subfigure[Player A aims at DB in the third throw when he has 50 remaining and B is also on 50.]
    {\label{figure:AndersonAspinallExample1}\includegraphics[width=0.48\linewidth]{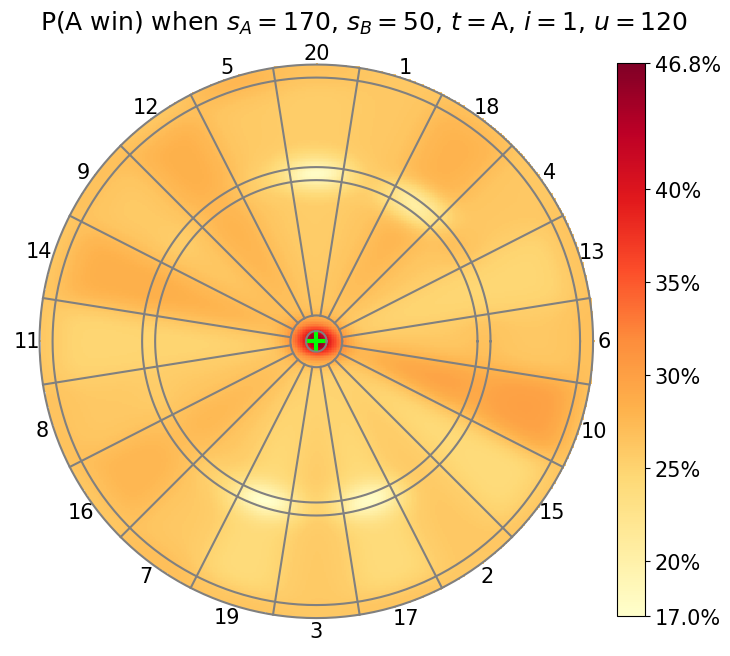}}
    \subfigure[Player A aims at S10 in the third throw when he has 50 remaining and B is on 150.]
    {\label{figure:AndersonAspinallExample2}\includegraphics[width=0.48\linewidth]{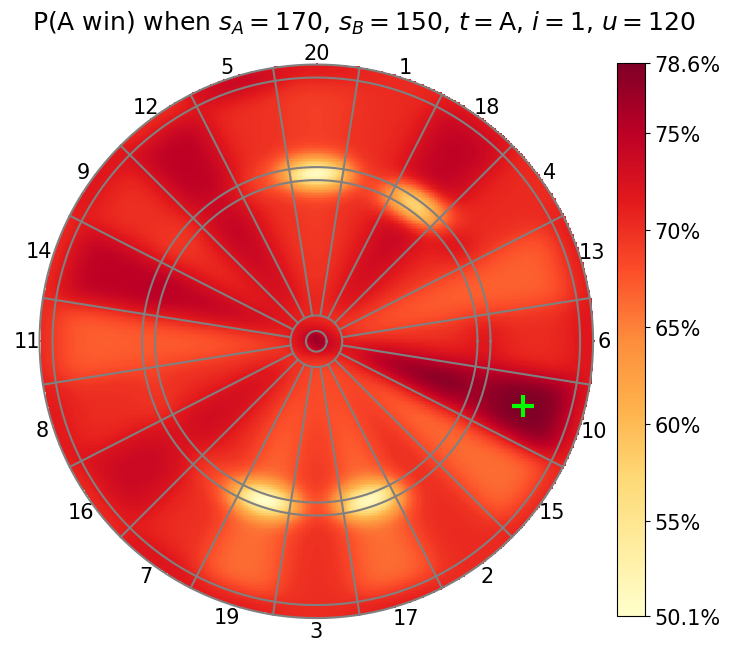}}
    \subfigure[$\piA^*$ and $\piNSA$ aim at DB.]
    {\label{figure:AndersonAspinallExample1-optloss}\includegraphics[width=0.48\linewidth]{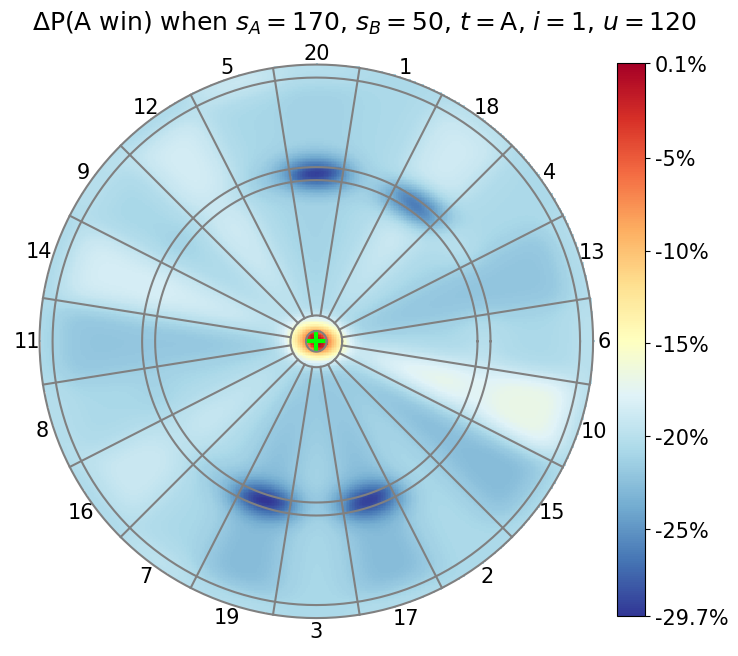}}
    \subfigure[$\piA^*$ aims at S10 and $\piNSA$ aims at DB.]
    {\label{figure:AndersonAspinallExample2-optloss}\includegraphics[width=0.48\linewidth]{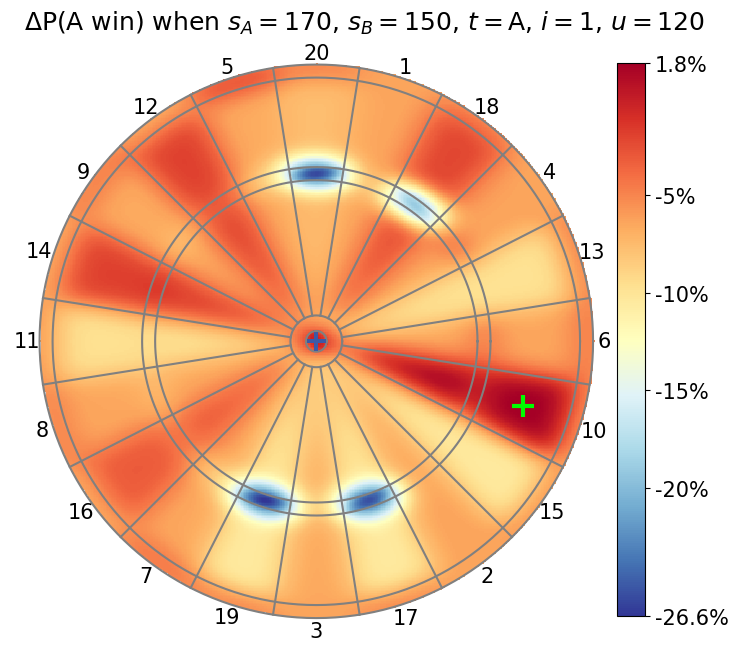}}
  \end{center}
  \small
  \emph{Note:} Figure (a) displays a heat-map of player A's options when it is his turn and player B only needs 50 to exit. His optimal target is the DB and every other target is considerably worse. In contrast, Player A is well ahead in Figure (b) and he will play defensively by aiming at S10 instead of DB on his third throw. Figures (c) and (d) display heat-maps $\Delta \Pb(\mbox{A win}) := Q^{\mbox{\tiny A-OPT}}(\bs, \ba) - \JpinsApibs(\bs)$ as a function of $\ba$ for the scenarios in (a) and (b), respectively. In (c) and (d), the blue and green crosses denote the targets chosen by $\piNSA$ and $\piA^*$, respectively.
  \caption{Hypothetical situations of Anderson (player A) vs. Aspinall (player B) with one throw remaining in Anderson's turn.}
  \label{figure:AndersonAspinallExample}
\end{figure}

Having solved for the equilibrium strategies of the ZSG, we can easily analyze hypothetical or real-world match-play strategic situations that might occur. These situations cannot be properly analyzed without solving the relevant ZSG as they are inherently strategic in that the optimal action for one player depends on the score of his opponent. Moreover, these kinds of situations are the most interesting and dramatic situations that arise in professional matches, since they occur near the end of a leg when both players are relatively close in score and therefore both have a reasonable chance of winning the leg.
In Figures \ref{figure:AndersonAspinallExample1} and \ref{figure:AndersonAspinallExample2} we consider two hypothetical situations of Anderson (player A) vs. Aspinall (player B),
where they both play their equilibrium strategies. The situation is identical in each figure except that Aspinall is on a score of 50 in the first and a score of 150 in the second.  Anderson has one throw remaining in his turn and needs $\sA - u = 170-120=50$ to exit on this throw. In Figure \ref{figure:AndersonAspinallExample1} Anderson must play aggressively and aim for the DB in which case he has a win-probability of 46.8\%. If he plays defensively and does not aim for the DB then his win-probability  will drop by between 15-20\% depending on where he aims. This is because Aspinall, on a score of 50, is very likely to exit on his next turn. In contrast, in Figure \ref{figure:AndersonAspinallExample2} we see that Anderson should play defensively and aim for S10 rather than DB. This is because Aspinall, on a score of 150, is far less likely to exit on his next turn. We note that Anderson's NS strategy would aim at DB in both scenarios. This hypothetical situation serves to emphasize the importance of playing strategically in the later part of a leg. The dependence of the optimal strategy on the opponent's score and his skill model also explains why it is difficult to explain the optimal strategy in terms of a few simple rules.

Figures \ref{figure:AndersonAspinallExample1-optloss} and \ref{figure:AndersonAspinallExample2-optloss} provide an alternative view on the two scenarios considered in Figures \ref{figure:AndersonAspinallExample1} and \ref{figure:AndersonAspinallExample2}, respectively. In order to understand them we first introduce some notation.
Let $\JpinsApibs(\bs)$ denote the probability of player A winning from state $\bs = (\sA,\sB,t,i,u)$ when player A uses $\piNSA$ and player B uses his equilibrium strategy $\piB^*$.
Let $Q^{\mbox{\tiny A-OPT}}(\bs, \ba)$ denote the probability of player A winning from state $\bs$ if he aims at $\ba$ on his current throw, plays his equilibrium strategy $\piA^*$ thereafter, and player B always plays $\piB^*$. We can therefore interpret $Q^{\mbox{\tiny A-OPT}}(\bs, \ba)$ as a so-called Q-value for player A. We let $\Delta \Pb(\mbox{A win}) := Q^{\mbox{\tiny A-OPT}}(\bs, \ba) - \JpinsApibs(\bs)$ which is a function of $\ba$. When $\ba$ is the target chosen by $\piA^*$ (denoted by the green crosses in Figures \ref{figure:AndersonAspinallExample1-optloss} and \ref{figure:AndersonAspinallExample2-optloss}), then $\Delta \Pb(\mbox{A win})$ is simply the difference in win-probabilities for player A playing $\piA^*$ instead of $\piNSA$ against player B playing $\piB^*$.

In Figure \ref{figure:AndersonAspinallExample1-optloss} the optimal $\ba$ is in the DB region and coincides with the action chosen by $\piNSA$. Hence the largest value of $\Delta \Pb(\mbox{A win})$ is very small and close to zero. (Note that $ \max_{\ba}\Delta \Pb(\mbox{A win}) \geq 0$ with equality occurring only if $\piA^* = \piNSA$ in state $\bs$ and all possible successor states to $\bs$.)
In contrast, we see from Figure \ref{figure:AndersonAspinallExample2-optloss} that the loss in win-probability if player A targets the DB (as recommended by $\piNSA$) rather than S10 (as recommended by $\piA^*$) is 1.8\% which is quite considerable.

We emphasize that this particular situation is not pathological in any sense. Indeed it is easy to find many situations, i.e. states $\bs$, where the difference in win-probabilities between following $\piNSA$ rather than $\piA^*$ is on the order of 1-2\%. These situations ultimately come down to $\piA^*$ accepting a slight increase in the expected number of turns to exit (which is what $\piNSA$ minimizes) for a slight increase in the probability of exiting in just one turn.

Finally, we remark that it would be disastrous for Anderson to aim at any of T17, T18, T19, or T20 since any of these scores would result in his going bust and therefore reverting to a score of 170 for his next turn. This would result in a massive drop in his win-probability and this is also clear from the heat-maps in Figure \ref{figure:AndersonAspinallExample}.

\section{Conclusions and Further Research}
\label{sec:Conclusions}

In this paper we formulated and solved the game of darts as a dynamic ZSG. We also estimated individual skill models for the top 16 professional players in the world during the 2019 season. Using the fitted skill models and our ZSG problem formulation and solution, we quantified the importance of playing strategically when computing an optimal strategy. For top professionals, we found that playing strategically (as opposed to non-strategically) results in an increase in win-probability of just 0.2\% - 0.6\% over a single leg but as much as 2.3\% over a best-of-35 legs match.

There are several directions for future research. First, if time-series data on the player's dart-throws was available then we could try to model the form of players as it varies through time.
Sufficiently granular data would also allow us to estimate a separate skill model for the first throw in a turn which is typically less accurate than the second and third throws. We discuss this further in Online Appendix \ref{sec:HotHand}. Following \citet{HotHands_RSSA}, it might also be of interest to model the within-turn hot-hand phenomenon and again, given sufficiently granular data, this would be easy to handle in our ZSG setting and we provide a formulation for this in Online Appendix \ref{sec:HotHand}. Modeling all of these features would therefore allow us to estimate a more accurate and state-dependent skill model for each player.

Given a more accurate skill-model, we could then use time-series data that included the dart-by-dart scoring evolution of matches, to estimate just how far the top players are deviating from their (estimated) optimal strategies.  For each player we could then estimate the loss in win-probability that results from playing sub-optimally. We note there are two possible reasons for playing sub-optimally: (i) a player not having an accurate assessment of his skill model and (ii) playing a sub-optimal strategy for his assumed skill strategy and we suspect both reasons lead to sub-optimal play on occasion. Even with the limitations of our data-set, it would be straightforward to quantify the loss in win-probability due to (i). In particular, we could for example assume skill model (\ref{eq:DM4}) was the ``true'' model while the player mistakenly assumes model (\ref{eq:DM2}) is correct. It would then be straightforward to assess how much is lost in win-probability while playing against other top players who do know their correct skill models and play their corresponding equilibrium strategies.

Another interesting direction for future work is in modifying the rules of darts so that the value of strategic playing, i.e. basing your strategy on your opponent's current score, becomes more important. TPT did some related analysis on this when they used a Metropolis-Hastings algorithm to search for configurations of the dartboard segments (1 to 20) that were more challenging for a darts player trying to maximize his expected score. Similarly, a relatively recent article  in \texttt{The Guardian} newspaper \citep{Guardian2013} discusses some other developments in this direction. Given the world-wide growth in the popularity of darts, we suspect variations on the game of 501 will be investigated and perhaps played in the future. In order to fully assess their impact on the professional game, skill models and the ability to solve the ZSG's will be required.

\section*{Acknowledgements}
We are very grateful to Christopher Kempf (\texttt{Twitter}  @ochepedia), Statistical Analyst for the PDC  for providing us with the data-set we used in Section \ref{sec:Numerics} and for some very insightful conversations. We are also grateful to the Area Editor, Nicola Secomandi, and the anonymous Associate Editor and referees for their helpful comments and suggestions. All errors are our own.

\renewcommand{\theequation}{\thesection-\arabic{equation}}
\setcounter{equation}{0}
\begin{appendices}

\section{Dartboard Geometry}
\label{sec:geom}

A standard competition dartboard has the following measurements (in millimeters):\\

\noindent
\begin{minipage}[b]{0.55\linewidth}
{ \begin{tabular}{l c }
  \toprule
  Distance & Measurement \\
  \midrule
  Center to DB wire & 6.35 \\
  Center to SB wire & 15.9 \\
  Center to inner triple wire & 99 \\
  Center to outer triple wire & 107 \\
  Center to inner double wire & 162 \\
  Center to outer double wire & 170 \\
  \bottomrule
\end{tabular}\vspace{.5cm}

Given these measurements and taking the center of DB to be $(0,0)$, it is straightforward to compute the map $z=g(x,y)$ that maps any $(x,y)$-coordinate on the dartboard to the corresponding dart score.
}
\end{minipage} 
\begin{minipage}[t]{0.45\linewidth}    \vspace{-6.0cm}
\begin{center}
\includegraphics[width=.75\linewidth]{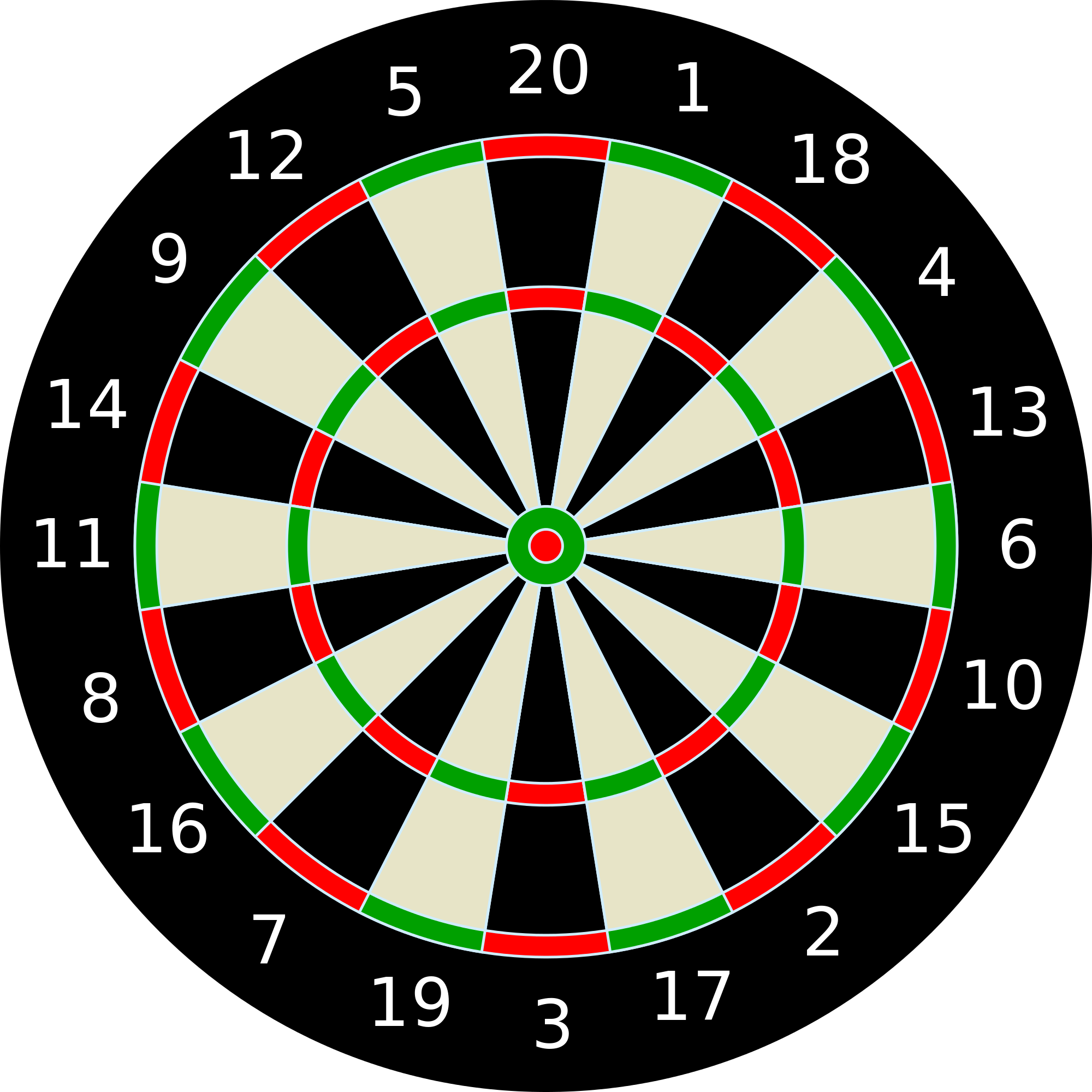}
\end{center}
\end{minipage}

\section{The Non-Strategic Model}
\label{sec:DP-formulation}

In this appendix we formulate the NS problem where a player's goal is simply to get to zero (by checking out on a double) as quickly as possible. We will formulate this as an MDP problem.

The state space ${\cal S}_{\mbox{\tiny NS}}$ for the MDP consists of all triples $\bs = (s,i,u)$ where $s$ is the score at the beginning of the turn, $i \in \{1,2,3\}$ is the number of throws remaining in the turn and $u$ is the score-to-date {\em within} the turn. We therefore have  $i=3$ and $u=0$ at the start of each turn.
The state transition function $f((s,i,u),z)$ depends on the current state $\bs$ and the realized dart score $z$. It is defined as follows.
\begin{equation}  \label{eq:Dynamics-DP_turn}
f((s,i,u),z) := \left\{
            \begin{array}{ll}
              (0,0,0), &  s = u + h(z) \mbox{ and } z \mbox{ a double, }  i\in\{1,2,3\}\\
              (s,3,0), & s - (u + h(z)) \leq 1 \\
              &  \hfill \mbox{ and not } (s=u+h(z), \ z \mbox{ a double}), i\in\{1,2,3\}\\
              (s,i-1,u+h(z)), & s > (u + h(z)) +1, \ i > 1 \\
              (s-u-h(z),3,0), & s > (u + h(z)) + 1, \ i = 1.
            \end{array}
          \right.
\end{equation}
The first two cases in (\ref{eq:Dynamics-DP_turn}) correspond to checking out and going bust, respectively. The third case corresponds to neither going bust nor checking out and still having at least one throw remaining in the turn. The final case corresponds to the last throw of a turn and neither checking out nor going bust.

We can model the per-stage costs of this MDP according to $c(s,i,u) := 1_{\{ i =3\}}$, 
so that we incur a cost of 1 at the beginning of each turn. Recall that the action-space ${\cal A}$ is the finite set of possible targets on the dartboard. It is state-independent.

An admissible policy $\pi := \{\pi_1, \pi_2, \ldots \}$ is a policy where each $\pi_k$ is a decision rule defining the target of the $k^{th}$ dart. That is, $\pi_k(\bs) \in {\cal A}$ for any $\bs \in {\cal S}_{\mbox{\tiny NS}}$ and with $k \in \mathbb{N}^+$ denoting the $k^{th}$ throw of the dart (across all turns). The total expected cost for any policy $\pi$ is then defined as
$V_{\pi}(\bs) := \lim_{K\to \infty} \Ex \left[ \sum_{k=1}^K c(\bs_k) \mid \bs_1 = \bs\right] $
where $K$ denotes the total number of permitted dart throws in a finite-horizon version of the problem.
Letting $\Pi$ represent the space of all admissible policies, the MDP problem is to solve for
\begin{equation} \label{eq:SSP-PRobForm1}
\VNST(\bs_1) := \min_{\pi \in \Pi } V_{\pi}(\bs_1)
\end{equation}
where $\bs_1 := (501,3,0)$ is the initial state of the game.
We recognize this infinite-horizon MDP as a stochastic shortest-path, i.e. SSP, problem where the goal is to check out in as few turns as possible.
A stationary policy $\pi$ is a policy where the decision rule $\pi_k$ does not depend on $k$. In this case we will use $\pi$ to denote both the policy and the decision rule $\pi(\bs)$. It is well known (see e.g. \citealp{BertVolII}) that there exists an optimal stationary policy under fairly modest assumptions.
One such assumption, which is sufficient but not necessary, is that every policy is {\em proper}, i.e. every policy will terminate with probability one. It is easy to argue that this assumption holds in our darts set-up.

A non-strategic model of a darts leg has previously been proposed and solved by \citet{Baird2020}. \citet{Kohler}  solved the same model except he ignored the turn feature. In particular, he assumed a player's score after going bust was his score immediately prior to the dart on which he went bust rather than his score at the beginning of the turn. (For the sake of completion we formulate the MDP solved by \citet{Kohler} in Online Appendix \ref{sec:IgnoringTurnDP}.)

\subsection{Solving the Non-Strategic Problem}
\label{sec:DP}

Chapter 2 of \citet{BertVolII} implies that the actions returned by the decision rules of an optimal stationary policy  for (\ref{eq:SSP-PRobForm1}) satisfy the Bellman equation
\begin{eqnarray}
\VNST(s,i,u) &=& \min_{\ba \in {\cal A}} \left\{\Ex[1_{\{i = 3\}} + \VNST(f((s,i,u),z)) \mid (s,i,u), \ba] \right\} \nonumber \\
&=& \min_{\ba \in {\cal A}} \left\{ \sum_{z} \left[1_{\{ i = 3\}} +\VNST(f((s,i,u),z))\right] \, p(z;\, \ba) \right\}  \label{eq:DP2}
\end{eqnarray}
for every feasible state $(s,i,u)$ with $\VNS(0,0,0):=0$.

As with our best-response problem (\ref{eq:DP-ABR1}), there is a monotonic structure to this DP (\ref{eq:DP2}): specifically, it is clear from the rules of the game, and indeed from the state transitions in (\ref{eq:Dynamics-DP_turn}), that the $s$ component of the state $(s,i,u)$ is non-increasing.
We can therefore use this to successively solve for $\VNST(2,i,u)$, $\VNST(3,i,u)$, $\ldots $, $\VNST(501,i,u)$ for all $i,u$.
If we let $\piNS$ denote the corresponding optimal strategy then $\piNS(s,i,u)$ is the optimal decision rule defining where the non-strategic (NS) dart thrower should aim when his current state is $(s,i,u)$. While the solution of SSP problems is still an active research area (e.g. see \citealp{Bert-Yu-2016}), our finite state and action spaces together with the assumption of policies being proper are sufficient to guarantee linear convergence of value iteration for solving (\ref{eq:DP2}). In particular, for a given error tolerance  $\epsilon$ under a suitably weighted norm (see e.g. \citealp{Bert-Yu-2016}), we can solve (\ref{eq:SSP-PRobForm1}) in $O(|{\cal S}_{\mbox{\tiny NS}}| \times |{\cal A}| \times \log (1/\epsilon))$ time. (We require $O(\log (1/\epsilon))$ value iterations and each iteration requires us to find the optimal action in each of $|{\cal S}_{\mbox{\tiny NS}}|$ states.  As there are $|{\cal A}|$ possible actions to consider for each state, each iteration therefore requires $O(|{\cal S}_{\mbox{\tiny NS}}| \times |{\cal A}|)$ time.) We need $O(\max \{|{\cal S}_{\mbox{\tiny NS}}|,\,  |{\cal A}|\})$ space for solving these problems.

In our numerical experiments, we took advantage of the aforementioned monotonicity structure and used policy iteration rather than value iteration for solving the SSPs. (While our running-time complexity was given for value iteration, as with the best-response problem for the ZSG, we again found policy iteration to be faster than value iteration.)

\subsection{Evaluating $\piNSA$ and $\piNSB$ Going Head-to-Head}
\label{sec:NS-Head-to-Head}

Suppose now that players A and B both play their NS strategies $\piNSA$ and $\piNSB$, respectively, as determined by (\ref{eq:DP2}). These strategies will be different in general as they will have different skill models and hence different $p(z;\, \ba)$ distributions.
It is straightforward to compute numerically the probability of player A winning a darts leg when both players play their respective NS strategies. This can be done recursively which provides an effective way of solving the linear system of equations to compute the probability of A winning before B.
The recursion is very similar to the recursion required for computing a player's best response (see (\ref{eq:DP-ABR1}) in Section \ref{sec:Solve-BRs}) except there is no maximization required as the strategies are fixed.

Even though the players are playing their NS strategies (which were computed using the state-space ${\cal S}_{\mbox{\tiny NS}}$), when we evaluate them going head-to-head we must use the state-space ${\cal S}_{\mbox{\tiny ZSG}}$ since the win-probabilities will be a function of the expanded state $\bs = (\sA,\sB,t,i,u) \in {\cal S}_{\mbox{\tiny ZSG}}$.  In order to compute these value functions, i.e. win-probabilities, we will need to compute $\pturn(s,s'; \pi)$, the probability of a player's score moving from  $s$ to  $s' \leq s$ over the course of a single turn when he plays strategy $\pi$.

It should be clear that exactly the same approach can be used to evaluate arbitrary strategies $\piA$ and $\piB$ (e.g. best-response, equilibrium strategies) going head-to-head against each other.

\end{appendices}

\bibliographystyle{ormsv080}
\bibliography{References_Darts}
\clearpage

\begin{appendices}
\renewcommand\appendixname{Online Appendix}

\begin{center}
\Large \bfseries Online Appendix: Play Like the Pros? Solving the Game of Darts as a Dynamic Zero-Sum Game
\end{center}

\section{Fitting the Skill Models via the EM Algorithm}
\label{app:EM}

We wish to fit each component of the normal skill model (\ref{eq:DM4}) for each of the 16 players. (Recall for each player we use six components in our model - one each for T20, T19, T18, T17, and DB, and one for the twenty doubles D1,...,D20 combined.) We will fix a particular component and let the data for this component be $\{(\ba_i,z_i,n_i )\}_{i=1}^p$.
The log-likelihood then satisfies
\begin{equation} \label{eq:CompleteLogLik1}
l(\bSigma; \, \{(\ba_i,z_i,n_i )\}_{i=1}^p ) \propto \sum_{i=1}^p n_i \, \log \left(\Pb \left((x,y) \in  R(z_i) \mid (x,y) \sim \mbox{N}_2\left(\ba_i, \bSigma\right)\right)\right)
\end{equation}
where $R(z_i)$ is the region defined by the dart $z_i$, i.e. $R(z_i) = \{(x,y) \in \mathbb{R}^2 \, : \, g(x,y) = z_i \}$, and $p$ is the number of regions in the data-set where darts landed. So for example, if we are fitting the skill model for any of the four treble components then $p=6$. Further details are provided in Section \ref{sec:Data}. For the DB component $p=22$ and for the remaining doubles component we have $p=20 \times 7 = 140$. Omitting its dependence on the data, we note that $l(\bSigma)$ can be easily computed via numerical integration but rather than attempting to maximize it w.r.t. $\bSigma$ directly, we can instead use the EM algorithm as suggested by TPT.

Let $n=n_1 + \cdots + n_p$ denote the total number of darts and let  $\bnu_i \in \mathbb{R}^2$ denote the unobserved realized position of the $i^{th}$ dart. The complete-data log-likelihood then satisfies
\begin{eqnarray}
\lcom(\bSigma; \, \{(\ba_i,\bnu_i,z_i,)\}_{i=1}^n) &=& -\frac{n}{2} \log | \bSigma | - \frac{1}{2} \sum_{i=1}^n \left(\ba_i - \bnu_i\right)^\top \bSigma^{-1} \left(\ba_i - \bnu_i \right) \nonumber \\
&=& -\frac{n}{2} \log | \bSigma | - \frac{1}{2} \mbox{tr} \left(\bSigma^{-1} \sum_{i=1}^n \left(\ba_i - \bnu_i\right) \left(\ba_i - \bnu_i \right)^\top \right)  \label{eq:LogLike1}
\end{eqnarray}
where $z_i = g(\bnu_i)$ for $i=1, \ldots , n$ and $\mbox{tr}(\cdot)$ denotes the trace operator.
(There is a slight abuse of notation here as $(\ba_i,z_i)$ now represents the target/outcome of the $i^{th}$ dart rather than the $i^{th}$ $(TR,z)$ combination in the data-set of which there are only $p$. The particular interpretation we have in mind should be clear from the context. For example, in going from (\ref{eq:M-step}) to (\ref{eq:M-step2}) we move from the former interpretation to the latter one.)
The E-step of the EM algorithm requires the calculation of
\begin{equation} \label{eq:E-step}
\Ex_{\bSigmaOld}\left[\lcom(\bSigma) \mid z \right] = -\frac{n}{2} \log | \bSigma | - \frac{1}{2} \mbox{tr} \left(\bSigma^{-1} \sum_{i=1}^n \Ex_{\bSigma_0}\left[ \left(\ba_i - \bnu_i\right) \left(\ba_i - \bnu_i \right)^\top \mid z_i\right]\right)
\end{equation}
where $\Ex_{\bSigmaOld}[\cdot \mid z]$ denotes the expectation operator conditional on the observed dart scores $z_1, \ldots , z_n$, and using the current parameter estimate $\bSigmaOld$. The M-step requires the maximization of (\ref{eq:E-step}) over $\bSigma$. This is identical to the usual maximum likelihood estimation for a multivariate Gaussian over an unknown covariance matrix and therefore has solution
\begin{eqnarray}
\bSigmaNew &=& \frac{1}{n} \sum_{i=1}^n \Ex_{\bSigmaOld}\left[ \left(\ba_i - \bnu_i\right) \left(\ba_i - \bnu_i \right)^\top \mid z_i\right] \label{eq:M-step} \\
&=& \frac{1}{n} \sum_{i=1}^{p} n_i \Ex_{\bSigmaOld}\left[ \left(\ba_i - \bnu_i\right) \left(\ba_i - \bnu_i \right)^\top \mid z_i\right]. \label{eq:M-step2}
\end{eqnarray}
The expectation in (\ref{eq:M-step2}) cannot be computed in closed form and so we follow the approach of TPT and compute them numerically via importance-sampling (IS). Specifically, we compute the expectation in (\ref{eq:M-step2}) as
\begin{equation} \label{eq:E-step2}
\Ex_{\bSigmaOld}\left[ \left(\ba_i - \bnu_i\right) \left(\ba_i - \bnu_i \right)^\top \mid z_i\right] \approx \sum_{j=1}^m w_{i,j} \left(\ba_i - \bnu_{i,j}\right) \left(\ba_i - \bnu_{i,j} \right)^\top
\end{equation}
where $m$ is the number of samples we used, the $\bnu_{i,j}$'s are samples of $\bnu_i$ and the $w_{i,j}$'s are corresponding IS weights. (We used $m=5,000$ when fitting our skill models.) Specifically
\begin{equation} \label{eq:IS1}
w_{i,j} = \frac{p_{i,j}/q_{i,j}}{\sum_{j=1}^m p_{i,j}/q_{i,j}}
\end{equation}
where $p_{i,j}$ is the density of $\bnu_{i,j}$ under the true distribution of $\bnu_{i}$ conditional on $z_i$ and using the parameter estimate $\bSigmaOld$, and $q_{i,j}$ is the sampling density of $\bnu_{i,j}$. Because we only know $p_{i,j}$ up to a multiplicative constant, we normalize in (\ref{eq:IS1}) to ensure the weights sum to 1. TPT suggested taking the sampling distribution $q_{i,j}$ to be uniform on $R(z_i)$ and this of course satisfies the IS requirement that $p=0$ whenever $q=0$. Moreover with this choice (\ref{eq:IS1}) is easy to compute since by assumption $q_{i,j} \propto 1$ and $p_{i,j}$ is the bivariate normal PDF with mean $\ba_i$ and variance-covariance matrix $\bSigmaOld$ evaluated at $\bnu_{i,j}$.
The EM algorithm then proceeds by iterating (\ref{eq:M-step2}) to (\ref{eq:IS1}) until convergence.

\section{Leg Win-Probability Results}
\label{sec:WinProbsByStratsLegs}

In this online appendix we display the win-probability results when each possible pair of opponents play the various strategies against each other.
Tables \ref{table:ProbabilityWinning1} and \ref{table:ProbabilityWinning2} display the leg win-probability for each player playing their non-strategic (NS), equilibrium, or BR (to NS) strategies against every other player.

\begin{table}[H]
\footnotesize
\centering
\captionsetup{justification=centering}
{
\renewcommand{\tabcolsep}{0.3mm}
\caption{\small Probability of Winning a Single Darts Leg}
\label{table:ProbabilityWinning1}
\begin{tabular}{l  c  cccccc  c  cccccc  c  cccccc  c  cccccc }

\toprule
\multirow{2}{*}{} & ~ & \multicolumn{6}{c}{p1 (Anderson)} & ~ & \multicolumn{6}{c}{p2 (Aspinall)} & ~ & \multicolumn{6}{c}{p3 (Chisnall)} & ~ & \multicolumn{6}{c}{p4 (Clayton)} \\
\cline{3-8} \cline{10-15} \cline{17-22} \cline{24-29}  \noalign{\smallskip}
~ & ~ & E-E & N-N & N-E & E-N & N-B & B-N & ~ & E-E & N-N & N-E & E-N & N-B & B-N & ~ & E-E & N-N & N-E & E-N & N-B & B-N & ~ & E-E & N-N & N-E & E-N & N-B & B-N \\
\midrule
p1&~&64.0&64.0&63.8&64.2&63.8&64.2&~&64.2&64.2&64.0&64.5&64.0&64.5&~&62.6&62.6&62.4&62.8&62.4&62.8&~&73.4&73.6&73.2&73.7&73.2&73.7 \\
p2&~&63.9&63.6&63.5&64.0&63.5&64.0&~&64.1&63.9&63.7&64.3&63.7&64.3&~&62.4&62.2&62.1&62.6&62.1&62.6&~&73.3&73.3&73.0&73.7&73.0&73.7 \\
p3&~&65.8&65.6&65.4&65.9&65.4&65.9&~&65.9&65.8&65.6&66.2&65.6&66.2&~&64.4&64.2&64.1&64.5&64.1&64.5&~&74.9&74.9&74.6&75.2&74.6&75.2 \\
p4&~&51.5&51.3&51.2&51.7&51.2&51.7&~&51.8&51.6&51.4&52.0&51.4&52.0&~&50.0&49.8&49.7&50.2&49.7&50.2&~&62.0&62.0&61.7&62.3&61.7&62.3 \\
p5&~&65.9&65.5&65.4&66.1&65.4&66.1&~&66.1&65.8&65.6&66.3&65.6&66.3&~&64.5&64.1&64.0&64.6&63.9&64.6&~&75.2&75.1&74.8&75.6&74.8&75.6 \\
p6&~&61.0&60.8&60.7&61.2&60.7&61.2&~&61.2&61.1&60.9&61.5&60.9&61.5&~&59.5&59.3&59.2&59.7&59.2&59.7&~&71.1&71.1&70.8&71.4&70.8&71.4 \\
p7&~&70.9&70.6&70.4&71.0&70.4&71.0&~&71.0&70.8&70.6&71.3&70.6&71.3&~&69.6&69.3&69.1&69.7&69.1&69.7&~&79.2&79.2&78.8&79.5&78.8&79.5 \\
p8&~&60.7&60.3&60.1&60.8&60.1&60.8&~&60.9&60.6&60.4&61.1&60.4&61.1&~&59.2&58.8&58.6&59.4&58.6&59.4&~&70.6&70.4&70.1&70.9&70.1&70.9 \\
p9&~&56.4&56.4&56.2&56.6&56.2&56.6&~&56.7&56.7&56.5&56.9&56.5&56.9&~&54.9&54.9&54.7&55.1&54.7&55.1&~&66.8&67.0&66.7&67.2&66.7&67.2 \\
p10&~&67.2&67.1&66.9&67.4&66.9&67.4&~&67.4&67.3&67.1&67.6&67.1&67.6&~&65.8&65.7&65.5&66.0&65.5&66.0&~&76.2&76.3&75.9&76.5&75.9&76.5 \\
p11&~&62.3&62.1&61.9&62.4&61.9&62.4&~&62.5&62.4&62.2&62.7&62.2&62.7&~&60.8&60.7&60.5&61.0&60.5&61.0&~&71.8&71.8&71.5&72.1&71.5&72.1 \\
p12&~&60.9&60.7&60.5&61.1&60.5&61.1&~&61.1&61.0&60.8&61.4&60.8&61.4&~&59.5&59.2&59.1&59.6&59.1&59.6&~&70.8&70.8&70.5&71.1&70.5&71.1 \\
p13&~&61.6&61.6&61.4&61.8&61.4&61.8&~&61.8&61.9&61.7&62.1&61.7&62.1&~&60.1&60.1&59.9&60.3&59.9&60.3&~&71.7&71.8&71.5&72.0&71.5&72.0 \\
p14&~&61.0&60.7&60.6&61.1&60.6&61.1&~&61.2&61.0&60.8&61.4&60.8&61.4&~&59.5&59.3&59.1&59.7&59.1&59.7&~&70.7&70.7&70.4&71.0&70.4&71.0 \\
p15&~&54.9&54.6&54.5&55.1&54.5&55.1&~&55.2&55.0&54.7&55.4&54.7&55.4&~&53.4&53.1&52.9&53.6&52.9&53.6&~&65.6&65.5&65.2&65.9&65.2&65.9 \\
p16&~&62.3&62.2&62.0&62.5&62.0&62.5&~&62.5&62.4&62.2&62.7&62.2&62.7&~&60.8&60.7&60.5&61.0&60.5&61.0&~&72.0&72.1&71.8&72.3&71.8&72.3 \\
\bottomrule
\\
\toprule
\multirow{2}{*}{} & ~ & \multicolumn{6}{c}{p5 (Cross)} & ~ & \multicolumn{6}{c}{p6 (Cullen)} & ~ & \multicolumn{6}{c}{p7 (van Gerwen)} & ~ & \multicolumn{6}{c}{p8 (Gurney)} \\
\cline{3-8} \cline{10-15} \cline{17-22} \cline{24-29}  \noalign{\smallskip}
~ & ~ & E-E & N-N & N-E & E-N & N-B & B-N & ~ & E-E & N-N & N-E & E-N & N-B & B-N & ~ & E-E & N-N & N-E & E-N & N-B & B-N & ~ & E-E & N-N & N-E & E-N & N-B & B-N \\
\midrule
p1&~&63.1&63.2&62.9&63.4&62.9&63.4&~&67.1&67.2&66.9&67.4&66.9&67.4&~&58.0&57.9&57.7&58.2&57.7&58.2&~&67.2&67.3&67.0&67.5&67.0&67.5 \\
p2&~&62.9&62.8&62.5&63.2&62.5&63.2&~&66.9&66.9&66.6&67.3&66.6&67.3&~&57.7&57.6&57.4&57.9&57.4&57.9&~&67.0&66.9&66.7&67.3&66.6&67.3 \\
p3&~&64.9&64.8&64.5&65.2&64.5&65.2&~&68.8&68.8&68.5&69.1&68.5&69.1&~&59.7&59.6&59.4&60.0&59.4&60.0&~&68.9&68.8&68.5&69.1&68.5&69.1 \\
p4&~&50.3&50.2&49.9&50.6&49.9&50.6&~&54.6&54.4&54.2&54.8&54.2&54.8&~&45.3&45.1&44.9&45.5&44.9&45.5&~&54.8&54.7&54.4&55.1&54.4&55.1 \\
p5&~&65.0&64.7&64.4&65.3&64.4&65.3&~&69.0&68.8&68.5&69.3&68.5&69.3&~&59.7&59.4&59.2&59.9&59.2&59.9&~&69.1&68.9&68.6&69.4&68.6&69.4 \\
p6&~&60.0&59.9&59.6&60.3&59.6&60.3&~&64.2&64.2&63.8&64.5&63.8&64.6&~&54.6&54.5&54.3&54.9&54.3&54.9&~&64.3&64.3&64.0&64.6&63.9&64.6 \\
p7&~&70.1&69.9&69.7&70.4&69.7&70.4&~&73.8&73.6&73.3&74.1&73.3&74.1&~&65.1&64.9&64.7&65.3&64.7&65.3&~&73.8&73.6&73.4&74.1&73.4&74.1 \\
p8&~&59.6&59.4&59.1&59.9&59.1&59.9&~&63.8&63.5&63.3&64.1&63.2&64.1&~&54.4&54.1&53.8&54.6&53.8&54.6&~&63.9&63.7&63.4&64.2&63.4&64.2 \\
p9&~&55.3&55.4&55.1&55.6&55.1&55.6&~&59.6&59.7&59.4&59.9&59.4&59.9&~&50.0&50.1&49.8&50.3&49.8&50.3&~&59.7&59.9&59.5&60.1&59.5&60.1 \\
p10&~&66.3&66.3&66.0&66.6&66.0&66.6&~&70.2&70.2&69.9&70.6&69.9&70.6&~&61.2&61.1&60.9&61.4&60.9&61.4&~&70.3&70.3&70.0&70.6&70.0&70.6 \\
p11&~&61.3&61.2&60.9&61.6&60.9&61.6&~&65.3&65.3&65.0&65.6&65.0&65.6&~&56.2&56.0&55.8&56.4&55.8&56.4&~&65.4&65.4&65.1&65.7&65.1&65.7 \\
p12&~&59.9&59.8&59.5&60.2&59.5&60.2&~&64.1&64.0&63.7&64.4&63.7&64.4&~&54.6&54.5&54.2&54.9&54.2&54.9&~&64.2&64.1&63.8&64.5&63.8&64.5 \\
p13&~&60.5&60.7&60.3&60.9&60.3&60.9&~&64.8&64.9&64.6&65.2&64.6&65.2&~&55.2&55.3&55.0&55.5&55.0&55.5&~&64.9&65.1&64.7&65.2&64.7&65.2 \\
p14&~&59.9&59.8&59.5&60.2&59.5&60.2&~&64.1&64.0&63.7&64.4&63.7&64.4&~&54.8&54.6&54.4&55.0&54.4&55.0&~&64.2&64.1&63.8&64.5&63.8&64.5 \\
p15&~&53.7&53.6&53.2&54.0&53.2&54.1&~&58.1&57.9&57.6&58.4&57.6&58.4&~&48.5&48.3&48.0&48.7&48.0&48.7&~&58.3&58.2&57.8&58.6&57.8&58.6 \\
p16&~&61.3&61.3&61.0&61.6&61.0&61.6&~&65.4&65.4&65.1&65.7&65.1&65.7&~&56.0&56.0&55.8&56.3&55.8&56.3&~&65.5&65.5&65.2&65.8&65.2&65.8 \\
\bottomrule
\end{tabular}}~\\
\justify
{\em Note:} Numbers are in percentages and the player from the left column throws first while the player (name in parentheses) from the top row throws second. N denotes the NS strategy, E denotes the equilibrium strategy, and B denotes the best-response strategy to an opponent's NS strategy.
We note that, to 1 decimal place, N-B and B-N are indistinguishable from N-E and E-N, respectively, in the vast majority of cases.
\end{table}

\begin{table}[H]
\footnotesize
\centering
\captionsetup{justification=centering}
{
\renewcommand{\tabcolsep}{0.25mm}
\caption{\small Probability of Winning a Single Darts Leg Continued}
\label{table:ProbabilityWinning2}
\begin{tabular}{l  c  cccccc  c  cccccc  c  cccccc  c  cccccc }

\toprule
\multirow{2}{*}{} & ~ & \multicolumn{6}{c}{p9 (Lewis)} & ~ & \multicolumn{6}{c}{p10 (Price)} & ~ & \multicolumn{6}{c}{p11 (Smith)} & ~ & \multicolumn{6}{c}{p12 (Suljovic)} \\
\cline{3-8} \cline{10-15} \cline{17-22} \cline{24-29}  \noalign{\smallskip}
~ & ~ & E-E & N-N & N-E & E-N & N-B & B-N & ~ & E-E & N-N & N-E & E-N & N-B & B-N & ~ & E-E & N-N & N-E & E-N & N-B & B-N & ~ & E-E & N-N & N-E & E-N & N-B & B-N \\
\midrule
p1&~&70.3&70.3&70.1&70.5&70.1&70.5&~&61.4&61.4&61.2&61.6&61.2&61.6&~&65.4&65.3&65.1&65.5&65.1&65.5&~&66.9&66.9&66.6&67.1&66.6&67.1 \\
p2&~&70.2&70.0&69.8&70.4&69.8&70.4&~&61.2&61.0&60.8&61.4&60.8&61.4&~&65.2&65.0&64.8&65.3&64.8&65.3&~&66.7&66.5&66.3&66.9&66.3&66.9 \\
p3&~&71.9&71.8&71.6&72.1&71.6&72.1&~&63.2&63.0&62.8&63.4&62.8&63.4&~&67.0&66.9&66.7&67.2&66.7&67.2&~&68.5&68.4&68.2&68.8&68.2&68.8 \\
p4&~&58.3&58.0&57.9&58.4&57.9&58.4&~&48.8&48.5&48.3&48.9&48.3&48.9&~&53.0&52.8&52.7&53.2&52.7&53.2&~&54.5&54.3&54.1&54.7&54.1&54.7 \\
p5&~&72.2&71.9&71.7&72.4&71.7&72.4&~&63.2&62.9&62.7&63.4&62.7&63.4&~&67.2&66.9&66.7&67.3&66.7&67.3&~&68.7&68.5&68.2&69.0&68.2&69.0 \\
p6&~&67.7&67.5&67.4&67.9&67.4&67.9&~&58.2&58.1&57.9&58.4&57.9&58.4&~&62.4&62.3&62.1&62.6&62.1&62.6&~&64.0&63.9&63.6&64.2&63.6&64.2 \\
p7&~&76.6&76.4&76.2&76.8&76.2&76.8&~&68.4&68.2&68.0&68.6&68.0&68.6&~&72.0&71.7&71.6&72.1&71.6&72.1&~&73.5&73.3&73.1&73.7&73.1&73.7 \\
p8&~&67.3&66.9&66.7&67.4&66.7&67.4&~&57.9&57.6&57.4&58.1&57.4&58.1&~&62.1&61.7&61.5&62.2&61.5&62.2&~&63.6&63.3&63.0&63.8&63.0&63.8 \\
p9&~&63.2&63.2&63.0&63.4&63.0&63.4&~&53.6&53.6&53.4&53.8&53.4&53.8&~&57.9&57.9&57.7&58.1&57.7&58.1&~&59.4&59.5&59.2&59.7&59.2&59.7 \\
p10&~&73.3&73.2&73.0&73.5&73.0&73.5&~&64.6&64.5&64.3&64.8&64.3&64.8&~&68.5&68.3&68.2&68.6&68.2&68.6&~&70.0&69.9&69.6&70.2&69.6&70.2 \\
p11&~&68.6&68.4&68.3&68.8&68.3&68.8&~&59.6&59.4&59.3&59.8&59.3&59.8&~&63.6&63.4&63.3&63.8&63.3&63.8&~&65.1&65.0&64.8&65.3&64.8&65.3 \\
p12&~&67.5&67.3&67.1&67.7&67.1&67.7&~&58.2&58.0&57.8&58.4&57.8&58.4&~&62.3&62.1&62.0&62.5&62.0&62.5&~&63.8&63.7&63.5&64.1&63.5&64.1 \\
p13&~&68.3&68.3&68.1&68.5&68.1&68.5&~&58.8&58.9&58.6&59.0&58.6&59.0&~&63.0&63.0&62.9&63.2&62.9&63.2&~&64.6&64.7&64.4&64.8&64.4&64.8 \\
p14&~&67.4&67.2&67.1&67.6&67.1&67.6&~&58.3&58.1&57.9&58.4&57.9&58.4&~&62.3&62.1&62.0&62.5&62.0&62.5&~&63.8&63.7&63.5&64.1&63.5&64.1 \\
p15&~&61.9&61.5&61.4&62.0&61.4&62.0&~&52.1&51.8&51.6&52.3&51.6&52.3&~&56.4&56.2&56.0&56.6&56.0&56.6&~&58.0&57.7&57.5&58.2&57.5&58.2 \\
p16&~&68.8&68.7&68.5&68.9&68.5&68.9&~&59.6&59.5&59.3&59.8&59.3&59.8&~&63.6&63.5&63.4&63.8&63.4&63.8&~&65.2&65.1&64.9&65.4&64.9&65.4 \\
\bottomrule
\\
\toprule
\multirow{2}{*}{} & ~ & \multicolumn{6}{c}{p13 (Wade)} & ~ & \multicolumn{6}{c}{p14 (White)} & ~ & \multicolumn{6}{c}{p15 (Whitlock)} & ~ & \multicolumn{6}{c}{p16 (Wright)} \\
\cline{3-8} \cline{10-15} \cline{17-22} \cline{24-29}  \noalign{\smallskip}
~ & ~ & E-E & N-N & N-E & E-N & N-B & B-N & ~ & E-E & N-N & N-E & E-N & N-B & B-N & ~ & E-E & N-N & N-E & E-N & N-B & B-N & ~ & E-E & N-N & N-E & E-N & N-B & B-N \\
\midrule
p1&~&66.6&66.5&66.4&66.7&66.3&66.7&~&66.7&66.7&66.4&66.9&66.4&66.9&~&71.5&71.5&71.3&71.7&71.3&71.7&~&65.8&65.7&65.5&65.9&65.5&65.9 \\
p2&~&66.4&66.1&66.0&66.6&66.0&66.6&~&66.5&66.3&66.1&66.7&66.1&66.7&~&71.4&71.3&71.0&71.7&71.0&71.7&~&65.6&65.4&65.2&65.8&65.2&65.8 \\
p3&~&68.3&68.0&67.9&68.4&67.9&68.4&~&68.3&68.2&68.0&68.5&68.0&68.5&~&73.0&73.0&72.7&73.3&72.7&73.3&~&67.5&67.3&67.1&67.6&67.1&67.6 \\
p4&~&54.0&53.7&53.6&54.1&53.6&54.1&~&54.3&54.2&53.9&54.5&53.9&54.5&~&59.5&59.4&59.2&59.8&59.2&59.8&~&53.3&53.0&52.9&53.4&52.9&53.4 \\
p5&~&68.5&68.1&68.0&68.6&68.0&68.6&~&68.5&68.2&68.0&68.7&68.0&68.7&~&73.4&73.2&72.9&73.6&72.9&73.6&~&67.6&67.3&67.1&67.8&67.1&67.8 \\
p6&~&63.7&63.4&63.3&63.8&63.3&63.8&~&63.8&63.7&63.4&64.0&63.4&64.0&~&68.9&68.9&68.6&69.3&68.6&69.3&~&62.8&62.6&62.5&63.0&62.5&63.0 \\
p7&~&73.3&73.0&72.9&73.4&72.9&73.4&~&73.3&73.0&72.8&73.5&72.8&73.5&~&77.6&77.5&77.2&77.9&77.2&77.9&~&72.5&72.2&72.1&72.7&72.1&72.7 \\
p8&~&63.3&62.8&62.7&63.4&62.7&63.4&~&63.4&63.1&62.9&63.6&62.9&63.6&~&68.5&68.2&67.9&68.8&67.9&68.8&~&62.4&62.0&61.9&62.6&61.9&62.6 \\
p9&~&59.0&58.9&58.8&59.2&58.8&59.2&~&59.2&59.3&59.1&59.5&59.1&59.5&~&64.5&64.6&64.3&64.8&64.3&64.8&~&58.2&58.2&58.0&58.4&58.0&58.4 \\
p10&~&69.7&69.5&69.4&69.8&69.4&69.8&~&69.7&69.7&69.4&70.0&69.4&70.0&~&74.4&74.4&74.1&74.7&74.1&74.7&~&68.9&68.7&68.6&69.1&68.6&69.1 \\
p11&~&64.8&64.5&64.5&64.9&64.5&64.9&~&64.9&64.8&64.6&65.1&64.6&65.1&~&69.8&69.7&69.5&70.1&69.5&70.1&~&64.0&63.8&63.7&64.2&63.7&64.2 \\
p12&~&63.5&63.2&63.1&63.7&63.1&63.7&~&63.6&63.5&63.3&63.9&63.3&63.9&~&68.7&68.6&68.4&69.0&68.4&69.0&~&62.7&62.5&62.3&62.9&62.3&62.9 \\
p13&~&64.3&64.2&64.1&64.4&64.1&64.4&~&64.4&64.4&64.2&64.6&64.2&64.6&~&69.5&69.7&69.4&69.9&69.4&69.9&~&63.4&63.4&63.2&63.6&63.2&63.6 \\
p14&~&63.5&63.2&63.1&63.6&63.1&63.6&~&63.6&63.5&63.3&63.9&63.3&63.9&~&68.6&68.5&68.3&68.9&68.3&68.9&~&62.7&62.5&62.3&62.9&62.3&62.9 \\
p15&~&57.6&57.2&57.1&57.7&57.1&57.7&~&57.8&57.6&57.3&58.0&57.3&58.0&~&63.2&63.0&62.7&63.5&62.7&63.5&~&56.7&56.4&56.3&56.9&56.3&56.9 \\
p16&~&64.9&64.7&64.6&65.0&64.6&65.0&~&65.0&64.9&64.7&65.2&64.7&65.2&~&70.0&70.0&69.7&70.3&69.7&70.3&~&64.1&63.9&63.8&64.2&63.8&64.2 \\
\bottomrule
\end{tabular}}~\\
\justify
{\em Note:} Numbers are in percentages and the player from the left column throws first while the player (name in parentheses) from the top row throws second. N denotes the NS strategy, E denotes the equilibrium strategy, and B denotes the best-response strategy to an opponent's NS strategy.
We note that, to 1 decimal place, N-B and B-N are indistinguishable from N-E and E-N, respectively, in the vast majority of cases.
\end{table}

\subsubsection*{Remark on Chisnall vs. Cross}
Near the end of Section \ref{sec:QuantStrat} we explained why the function $\mbox{Gain}(\piA^*,\piNSA,N)$ typically monotonically increases in $N$ until some value $N^*$ say, after which it decreases monotonically to zero. An exception to this general rule occurs when playing $\piNSA$ makes A an underdog but switching to playing $\piA^*$ makes him a favorite. In this latter case we would see $\mbox{Gain}(\piA^*,\piNSA,N)$ monotonically increases to 1 as $N \to \infty$ and we observed this in the case of Chisnall (player A) vs. Cross (player B) who are very evenly matched. It may be seen from Table \ref{table:ProbabilityWinning1} that when Chisnall and Cross both play their equilibrium strategies then Chisnall has a slight edge as he wins 64.9\% of the legs he starts whereas Cross wins 64.5\% of the legs he starts. In contrast, if Chisnall plays his NS strategy then Cross will win 64.6\% of the legs he starts whereas Chisnall will win 64.5\% of the legs he starts thereby giving Cross an edge.
This then is a case where (at least in the limit of large $N$) taking account of an opponent's score and playing the equilibrium strategy makes all the difference between winning and losing.

\section{Computing Match-Win Probabilities}
\label{app:MatchWinProbs}

In this online appendix we describe the calculations that allow us to convert the win-probabilities for single legs into the win-probability for a match consisting of several legs.
Following on from Section \ref{sec:QuantStrat}, we wish to compute $\mbox{Gain}(\piA^*,\piNSA,N)$ as defined in (\ref{eq:Adv1}) and which we repeat here for the sake of convenience:
\begin{equation} \label{eq:Adv111}
\mbox{Gain}(\piA^*,\piNSA,N) := \PA(\piA^*,N) - \PA(\piNSA,N).
\end{equation}
$\mbox{Gain}(\piA^*,\piNSA,N)$ is therefore player A's increase in probability of winning a match of $N$ legs from playing $\piA^*$ rather than $\piNSA$ against player B playing $\piB^*$. Each term in (\ref{eq:Adv111}) is easily calculated.
Specifically, let $V$ denote the number of legs that A wins out of $N$ legs in total where A starts the first leg and the players then alternate in starting legs.
We let $\VVA$ and $\VVB$ denote the number of legs won by A that were started by A and B, respectively, and then we have $V = \VVA+\VVB$.
As a typical darts match has an odd number of legs, we will assume $N=2K+1$ so that a player needs to win at least $K+1$ legs to win the match. Therefore, we have
\begin{eqnarray}
\PA(\piA^*,N) &=& \Pb(V \geq K+1) \nonumber \\
&=& \sum_{j=1}^{K+1} \Pb(V \geq K+1 \mid \VVA =j ) \Pb(\VVA = j) \nonumber \\
&=& \sum_{j=1}^{K+1} \Pb(\VVB \geq K+1-j  ) \Pb(\VVA = j)  \label{eq:Adv2}
\end{eqnarray}
which is easily calculated since $\VVA \sim \mbox{Bin}(K+1,\pA(\piA^*))$ and $\VVB \sim \mbox{Bin}(K,\pB(\piA^*))$ where 
\begin{eqnarray*}
\pA(\piA) &:=& \Pb(\mbox{A wins leg} \mid \mbox{A starts leg, A plays } \piA) \label{eq:WinProbMatch123456} \\
\pB(\piA) &:=& \Pb(\mbox{A wins leg} \mid \mbox{B starts leg, A plays } \piA). \label{eq:WinProbMatch123457}
\end{eqnarray*}
$\PA(\piNSA,N)$ is computed as in (\ref{eq:Adv2}) except in this case $\VVA \sim \mbox{Bin}(K+1,\pA(\piNSA))$ and $\VVB \sim \mbox{Bin}(K,\pB(\piNSA))$.

\subsection{Further Results for $\mbox{Gain}(\piA^*,\piNSA,N)$}
Tables \ref{table:Match_leg1}, \ref{table:Match_leg21} and \ref{table:Match_leg35} display values for  $\mbox{Gain}(\piA^*,\piNSA,N)$ when $N=1$, $21$ and $35$, respectively.

\begin{table}[ht]
\small
\centering
\captionsetup{justification=centering}
{
\renewcommand{\tabcolsep}{1mm}
\caption{\small $\mbox{Gain}(\piA^*,\piNSA,N)$ in a Match of $N=1$ Leg}
\label{table:Match_leg1}
\begin{tabular}{l  cccc cccc cccc cccc}
\toprule
~ & p1 & p2 & p3 & p4 & p5 & p6 & p7 & p8 & p9 & p10 & p11 & p12 & p13 & p14 & p15 & p16 \\
\midrule
p1&0.2&0.2&0.2&0.2&0.2&0.2&0.2&0.2&0.2&0.2&0.2&0.2&0.2&0.2&0.2&0.2 \\
p2&0.4&0.4&0.4&0.3&0.4&0.4&0.3&0.4&0.4&0.4&0.3&0.4&0.4&0.4&0.4&0.4 \\
p3&0.3&0.3&0.3&0.3&0.3&0.3&0.3&0.3&0.3&0.3&0.3&0.3&0.4&0.3&0.3&0.3 \\
p4&0.4&0.4&0.4&0.3&0.4&0.4&0.4&0.4&0.4&0.4&0.4&0.4&0.4&0.4&0.4&0.4 \\
p5&0.5&0.5&0.5&0.4&0.5&0.5&0.5&0.5&0.5&0.5&0.5&0.5&0.5&0.5&0.5&0.5 \\
p6&0.4&0.4&0.4&0.3&0.4&0.4&0.4&0.4&0.4&0.4&0.3&0.4&0.4&0.4&0.4&0.4 \\
p7&0.4&0.4&0.4&0.4&0.5&0.5&0.4&0.4&0.4&0.5&0.4&0.4&0.5&0.4&0.4&0.4 \\
p8&0.5&0.5&0.5&0.5&0.6&0.6&0.5&0.5&0.5&0.6&0.5&0.5&0.6&0.5&0.5&0.6 \\
p9&0.2&0.2&0.2&0.2&0.2&0.2&0.2&0.2&0.2&0.2&0.2&0.2&0.2&0.2&0.2&0.2 \\
p10&0.3&0.3&0.3&0.3&0.3&0.3&0.3&0.3&0.3&0.3&0.3&0.3&0.3&0.3&0.3&0.3 \\
p11&0.3&0.3&0.3&0.3&0.4&0.4&0.3&0.3&0.3&0.3&0.3&0.3&0.4&0.3&0.3&0.3 \\
p12&0.4&0.4&0.4&0.3&0.4&0.4&0.4&0.4&0.4&0.4&0.4&0.4&0.4&0.4&0.4&0.4 \\
p13&0.2&0.2&0.2&0.2&0.2&0.2&0.2&0.2&0.2&0.2&0.2&0.2&0.2&0.2&0.2&0.2 \\
p14&0.4&0.4&0.4&0.3&0.4&0.4&0.4&0.4&0.4&0.4&0.4&0.4&0.4&0.4&0.4&0.4 \\
p15&0.5&0.4&0.5&0.4&0.5&0.5&0.5&0.5&0.5&0.5&0.4&0.5&0.5&0.4&0.4&0.5 \\
p16&0.3&0.3&0.3&0.2&0.3&0.3&0.3&0.3&0.3&0.3&0.3&0.3&0.3&0.3&0.3&0.3 \\
\bottomrule
\end{tabular}}~\\
\justify
{\em Note:} Numbers are in percentages. The player on the left column is player A who begins the first leg of the match. The player on the top row is player B who uses his equilibrium strategy $\piB^*$ throughout.
\end{table}

\begin{table}[H]
\small
\centering
\captionsetup{justification=centering}
{
\renewcommand{\tabcolsep}{1mm}
\caption{\small $\mbox{Gain}(\piA^*,\piNSA,N)$ in a Match of $N=21$ Legs}
\label{table:Match_leg21}
\begin{tabular}{l  cccc cccc cccc cccc}
\toprule
~ & p1 & p2 & p3 & p4 & p5 & p6 & p7 & p8 & p9 & p10 & p11 & p12 & p13 & p14 & p15 & p16 \\
\midrule
p1&0.8&0.8&0.8&0.4&0.8&0.8&0.7&0.7&0.6&0.8&0.7&0.7&0.8&0.7&0.5&0.8 \\
p2&1.2&1.1&1.1&0.6&1.2&1.2&1.0&1.1&1.0&1.2&1.1&1.1&1.3&1.1&0.9&1.2 \\
p3&0.9&0.9&0.9&0.4&1.0&0.9&0.8&0.9&0.7&1.0&0.8&0.9&1.0&0.8&0.6&0.9 \\
p4&0.8&0.8&0.7&1.1&0.8&1.1&0.4&1.1&1.3&0.6&0.9&1.0&1.1&1.0&1.3&1.0 \\
p5&1.5&1.5&1.6&0.7&1.6&1.5&1.4&1.4&1.1&1.6&1.4&1.4&1.6&1.4&1.0&1.5 \\
p6&1.3&1.3&1.2&0.8&1.3&1.4&1.0&1.3&1.2&1.2&1.2&1.3&1.4&1.3&1.1&1.4 \\
p7&1.0&1.0&1.1&0.3&1.2&0.9&1.2&0.8&0.6&1.2&0.9&0.9&1.0&0.8&0.5&1.0 \\
p8&1.5&1.5&1.5&1.1&1.5&1.7&1.0&1.6&1.6&1.4&1.5&1.7&1.8&1.6&1.5&1.7 \\
p9&0.6&0.6&0.5&0.5&0.5&0.7&0.3&0.6&0.7&0.5&0.6&0.6&0.7&0.6&0.7&0.6 \\
p10&0.9&0.9&1.0&0.3&1.0&0.8&0.9&0.8&0.6&1.0&0.8&0.8&0.9&0.8&0.5&0.9 \\
p11&0.9&0.9&0.9&0.6&1.0&1.0&0.7&0.9&0.9&0.9&0.9&1.0&1.1&0.9&0.8&1.0 \\
p12&1.1&1.1&1.1&0.8&1.2&1.2&0.8&1.2&1.1&1.1&1.1&1.2&1.3&1.1&1.0&1.2 \\
p13&0.6&0.6&0.5&0.4&0.6&0.6&0.4&0.6&0.5&0.5&0.5&0.6&0.6&0.6&0.5&0.6 \\
p14&1.1&1.1&1.1&0.8&1.2&1.2&0.8&1.2&1.1&1.1&1.1&1.2&1.3&1.1&1.0&1.2 \\
p15&1.1&1.1&1.0&1.2&1.0&1.4&0.6&1.3&1.5&0.9&1.1&1.3&1.4&1.2&1.5&1.2 \\
p16&0.8&0.8&0.8&0.5&0.9&0.9&0.6&0.8&0.8&0.8&0.8&0.9&1.0&0.8&0.7&0.9 \\
\bottomrule
\end{tabular}}~\\
\justify
{\em Note:} Numbers are in percentages. The player on the left column is player A who begins the first leg of the match. The player on the top row is player B who uses his equilibrium strategy $\piB^*$ throughout.
\end{table}

\begin{table}[ht]
\small
\centering
\captionsetup{justification=centering}
{
\renewcommand{\tabcolsep}{1mm}
\caption{\small $\mbox{Gain}(\piA^*,\piNSA,N)$ in a Match of $N=35$ Legs}
\label{table:Match_leg35}
\begin{tabular}{l  cccc cccc cccc cccc}
\toprule
~ & p1 & p2 & p3 & p4 & p5 & p6 & p7 & p8 & p9 & p10 & p11 & p12 & p13 & p14 & p15 & p16 \\
\midrule
p1&1.0&1.0&1.0&0.3&1.1&1.0&0.8&0.9&0.7&1.0&0.9&0.9&1.0&0.9&0.6&1.0 \\
p2&1.5&1.5&1.4&0.5&1.5&1.5&1.1&1.4&1.1&1.4&1.4&1.4&1.6&1.4&0.9&1.5 \\
p3&1.2&1.2&1.2&0.3&1.3&1.1&1.0&1.0&0.7&1.2&1.1&1.1&1.2&1.0&0.6&1.1 \\
p4&0.7&0.7&0.6&1.5&0.6&1.1&0.2&1.1&1.5&0.5&0.9&1.1&1.1&1.0&1.6&0.9 \\
p5&1.9&1.9&2.0&0.5&2.1&1.8&1.6&1.7&1.2&2.0&1.7&1.7&1.9&1.7&1.0&1.9 \\
p6&1.6&1.6&1.5&0.8&1.6&1.8&0.9&1.7&1.5&1.4&1.6&1.7&1.9&1.6&1.3&1.7 \\
p7&1.2&1.1&1.3&0.2&1.4&0.9&1.5&0.8&0.4&1.5&0.9&0.8&1.0&0.8&0.3&1.0 \\
p8&1.9&1.9&1.7&1.2&1.8&2.2&1.0&2.1&2.0&1.6&1.9&2.1&2.3&2.0&1.8&2.1 \\
p9&0.6&0.6&0.5&0.6&0.6&0.8&0.2&0.8&0.9&0.5&0.7&0.8&0.8&0.7&0.8&0.7 \\
p10&1.2&1.1&1.2&0.2&1.3&1.0&1.1&0.9&0.6&1.3&1.0&1.0&1.1&0.9&0.5&1.1 \\
p11&1.2&1.2&1.1&0.6&1.2&1.3&0.7&1.2&1.0&1.1&1.1&1.2&1.4&1.2&0.9&1.3 \\
p12&1.4&1.4&1.3&0.8&1.4&1.6&0.8&1.5&1.4&1.2&1.4&1.5&1.7&1.5&1.2&1.5 \\
p13&0.7&0.7&0.6&0.4&0.7&0.8&0.4&0.7&0.6&0.6&0.7&0.7&0.8&0.7&0.6&0.8 \\
p14&1.4&1.4&1.3&0.8&1.4&1.6&0.8&1.5&1.4&1.2&1.4&1.5&1.7&1.5&1.2&1.5 \\
p15&1.1&1.1&0.9&1.5&1.0&1.6&0.4&1.5&1.9&0.8&1.2&1.5&1.6&1.4&1.9&1.4 \\
p16&1.1&1.1&1.0&0.5&1.1&1.1&0.7&1.1&0.9&1.0&1.0&1.1&1.2&1.0&0.8&1.1 \\
\bottomrule
\end{tabular}}~\\
\justify
{\em Note:} Numbers are in percentages. The player on the left column is player A who begins the first leg of the match. The player on the top row is player B who uses his equilibrium strategy $\piB^*$ throughout.
\end{table}

\section{Other Problem Formulations}
\label{sec:OtherProbForms}

In this online appendix we describe some additional problem formulations including a formulation where we model the hot-hands phenomenon within a turn.
In Online Appendix \ref{sec:IgnoringTurnDP} we consider the non-strategic DP formulation when we ignore the turn feature. In Online Appendix \ref{sec:HotHand} we briefly explain how we could easily account for the hot-hands phenomenon {\em within} a turn. 

\subsection{Ignoring the ``Turn'' Feature in the DP Formulation}
\label{sec:IgnoringTurnDP}

An interesting problem in its own right is the problem whereby a player aims to minimize the expected number of throws to get to zero. Such a player pays no attention to the score of his opponent and is therefore non-strategic. To the best of our knowledge \cite{Kohler} was the first to solve this problem.
(He used the skill model $(x,y)  \sim  \mbox{N}_2(\ba,\sigma^2 \bI)$  and due to limited computing power at the time, used various numerical approximations together with a branch-and-bound technique to solve the DP.)

The state space for the DP is $(s,i)$ where $s \in \{0, 2, \ldots, 501\}$ denotes the player's current score and $i$ denotes how many throws have already been taken.
The state transition function $f(s_i,z)$ depends on the current state $s_i$ and the realized dart score. It is defined as follows.
\begin{equation} \label{eq:State_Dynamics}
        s_{i+1} = f(s_i,z) := \left\{
                    \begin{array}{ll}
                      s_i - h(z), & \ s_i - h(z) > 1  \\
                        0, & \ h(z)=s_i \mbox{ and } z \mbox{ a double} \\
                      s_i, & \ \hbox{otherwise}
                    \end{array}
                  \right.
\end{equation}
where $h(z) \geq 0$ is the numerical score when the $(i+1)^{st}$ throw achieves score $z$.
The three cases in \eqref{eq:State_Dynamics} correspond to a valid throw, checking out, and going bust, respectively.
Let $\VNS(s,i)$ denote the value function, and then the Bellman equation satisfies
\begin{eqnarray} \label{eq:DP-Bellman}
\VNS(s_i,i) &=& \min_{\ba \in {\cal A}} \Ex \left[ V(s_{i+1},i+1) \mid (s_i,i), \ba \right] \nonumber \\
&=& \min_{\ba \in {\cal A}} \, \sum_{z} \, \VNS(f(s_i,z) ,i+1) \, p(z;\, \ba)
\end{eqnarray}
with the terminal condition $\VNS(0,i) = i$.
In this non-strategic version of the game we want to compute $\VNS(501,0)$.

As with the DP in Appendix \ref{sec:DP}, there is a monotonic structure to this DP. Specifically, the state $s_i$ is non-increasing in $i$ so that $s_{i+1} \leq s_i$ for all $i$. We can take advantage of this by redefining $\VNS(s)$ to be the minimum expected number of throws required to reach zero starting from state $s \geq 2$. We then have
\begin{eqnarray}
\VNS(s) &=& \min_{\ba \in {\cal A}} \left\{1 + \Ex[\VNS(f(s,z)) \mid s, \ba] \right\} \nonumber \\
&=& \min_{\ba \in {\cal A}} \left\{1 + \sum_{z} \VNS(f(s,z)) \, p(z;\, \ba) \right\}  \label{eq:DP1}
\end{eqnarray}
with $\VNS(0) :=0$. We can solve (\ref{eq:DP1}) successively for $s=2, \ldots , 501$.

\subsection{Modeling a Within-Turn ``Hot-Hand'' Phenomenon}
\label{sec:HotHand}

Given a sufficiently rich data-set, one could easily estimate a different skill model $p_i(x,y; \ba)$ for each throw $i$ within a turn. Indeed it is well known, see e.g. \citet{HotHands_RSSA}, that the success rates for the second and third throws within a turn are considerably higher than the success rate for the first throw. An obvious and widely accepted explanation for this is that errors in the first throw can be used to re-calibrate and hence improve the accuracy of the second and third throws.

\citet{HotHands_RSSA} also provide some evidence for the hot-hands phenomenon within a turn but find little evidence for it across turns. Here we simply observe that it would be straightforward to model the within-turn hot-hands phenomenon by a simple and small expansion of the state space. Consider the DP formulation in Appendix \ref{sec:DP}, for example. Recall the state space for this DP consisted of $(s,i,u)$ where $s$ is the score at the beginning of the turn, $i \in \{1,2,3\}$ is the number of throws remaining in the turn and $u$ is the score-to-date {\em within} the turn. To model the hot-hands phenomenon within a turn, we instead use state vectors $(s,i,u)$ where $s$ and $u$ are as before but $i \in \{1,\ldots, 7\}$ used to encode the number of throws remaining in the turn (as before) as well as the success(es) of any previous throw(s) within the turn. For example, we could use $i=1$ to represent that there are 3 throws remaining in the turn. Then $i=2$ and $i=3$ could denote the states where there are 2 remaining throws in the turn and the first throw was a success or failure, respectively. Finally, we could use $i=4,\ldots , 7$ to denote the states where there is just 1 throw remaining. There are 4 such states depending on the success or failure outcomes of the first 2 throws.

This formulation requires slightly more than doubling the number of states and it clearly also applies to the BR and ZSG problem formulations. It would therefore only require a modest increase in computational work to solve for these problems. The main difficulty of course would be in obtaining a sufficiently rich data-set to estimate the probability distributions required for the state transitions.

\section{Speeding Up Solving the ZSG}
\label{app:SpeedUps}

As noted at the beginning of Section \ref{sec:RunTimes}, we were able to obtain significant improvements in the performance of Algorithm \ref{alg:ZSG} through the use of vectorization that takes advantage of GPUs. Beyond standard uses of vectorization (e.g. evaluating all actions for a state in a vectorized operation, which we also implemented), we found the following use of vectorization to be particularly efficient.

The core step of Algorithm \ref{alg:ZSG} is in applying Algorithm \ref{alg:BR} for solving the best-response problem in the while loop. Consider then this step where we implement Algorithm \ref{alg:BR} using policy iteration for a given $(\sA,\sB)$. We need to update $\VA(\sA,\sB,i,u)$ via \eqref{eq:DP-ABR1} for all $(i, u)$ in the policy improvement step.
It is clear from the state transition dynamics in \eqref{eq:Dynamics-DP-ABR} that the value of $i$ must change during each state transition, i.e., the state $(\sA,\sB,i,u)$ transitions to either the next throw, the next turn, going bust, or the end of the game, all of which result in a change in the value of $i$.
Therefore, updating $\VA(\sA,\sB,i,u)$ in the policy improvement step does not require $\VA(\sA,\sB,i,u')$ for any states $(\sA,\sB,i,u')$. This means that $\VA(\sA,\sB,i,u)$ can be updated simultaneously for states with the same value of $i$.
Specifically, given $(\sA,\sB)$, we update $\VA(\sA,\sB,i=1,u)$ for all $u$ in one operation, then update $\VA(\sA,\sB,i=2,u)$ for all $u$ in another operation, and finally update $\VA(\sA,\sB,i=3,u)$.
To achieve this, we load all the information required into several large matrices and then evaluate the Bellman equation \eqref{eq:DP-ABR1} using a GPU. GPUs of course are extremely efficient in executing large matrix algebra.

\end{appendices}

\end{document}